\documentclass[structabstract]{aa}  
\usepackage{graphicx}
\usepackage{txfonts}
\usepackage{longtable}

\begin{document}
   \title{The atmospheric structure and fundamental parameters of the 
red supergiants AH Sco, UY Sct, and KW Sgr}

   \titlerunning{The atmospheric structure and fundamental parameters 
of three red supergiants}

   \subtitle{}

   \author{B. Arroyo-Torres \inst{1} \and
          M. Wittkowski \inst{2} \and J. M. Marcaide \inst{1} 
\and P. H. Hauschildt \inst{3}
          }

   \institute{Dpt. Astronomia i Astrof\' isica, Universitat de Val\`encia, 
C/Dr. Moliner 50, 46100, Burjassot, Spain \email{belen.arroyo@uv.es}
         \and ESO, Karl-Schwarzschild-St. 2, 85748, Garching bei M\" unchen, 
Germany
         \and Hamburger Sternwarte, Gojenbergsweg 112, 21029, Hamburg, Germany
             }

   \date{Received 14 December 2012 ; accepted 18 April 2013}

 
  \abstract
   {}
   {We present the atmospheric structure and the fundamental properties 
of the red supergiants (RSGs) AH~Sco, UY~Sct, and KW~Sgr based on 
VLTI/AMBER observations.}
   {We carried out spectro-interferometric observations of AH~Sco, 
UY~Sct, and KW~Sgr in the near-infrared K band (1.92-2.47\,$\mu$m) with the 
VLTI/AMBER instrument with spatial and spectral resolutions of 
3 milliarcseconds and 1500, respectively, and compared the data
to a new grid of hydrostatic PHOENIX model atmospheres.}
   {In our visibility data, we observe molecular layers 
of water and CO in extended atmospheres. For a uniform disk modeling, 
we observe size increases at the water band centered at 1.9\,$\mu$m 
of 10\% to 25\% and at the CO bandheads at 2.3-2.5\,$\mu$m of 20\%-35\%
with respect to the near-continuum bandpass at around 2.20\,$\mu$m. 
Our near-infrared spectra of AH~Sco, UY~Sct, 
and KW~Sgr are well reproduced by the PHOENIX model atmospheres. 
The continuum visibility values are consistent
with a limb-darkened disk as predicted by the PHOENIX models.
However, the model visibilities do not predict the large 
observed 
extensions of the molecular layers.
Comparing the continuum visibility values to PHOENIX models,
we estimate the Rosseland-mean photospheric angular 
diameters of AH~Sco, UY~Sct, and KW~Sgr to be 5.81$\pm$0.15~mas, 
5.48$\pm$0.10~mas, and 3.91$\pm$0.25~mas, respectively. 
Together with the distance and the spectro-photometry, we calculate
radii of 1411$\pm$124\,R$_{\odot}$ for AH~Sco, 1708$\pm$192\,R$_{\odot}$ 
for UY~Sct, and 1009$\pm$142\,R$_{\odot}$ for KW~Sgr and effective 
temperatures of 3682$\pm$190\,K for AH~Sco, 3365$\pm$134\,K for UY~Sct, 
and 3720$\pm$183\,K for KW~Sgr.}
   {AH~Sco, UY~Sct, and KW~Sgr exhibit extended atmospheric layers of 
H$_{2}$O and CO. 
The PHOENIX atmosphere models predict the spectra and the continuum
visibility values, but cannot reproduce the large extensions
of the molecular layers. This indicates that the opacities of
the molecular bands are included, but that the model atmospheres
are too compact compared to the observations. The observed extended
layers may be levitated by processes such as pulsation or convection,
which are not included in the hydrostatic atmospheric models.
The location of the targets in the HR-diagram is confirmed to be close to, and possibly slightly to the right of, the Hayashi limit of recent evolutionary tracks corresponding to masses between about 20\,M$_\odot$ and 40\,M$_\odot$.
}
   \keywords{supergiants -- Star: fundamental parameters --
                Star: atmospheres -- Hertzsprung-Russell and C-M diagrams -- 
Star: individual: AH Sco, UY Sct and KW Sgr.
               }

\maketitle

%

\section{Introduction}
Our research aims to investigate the circumstellar environment 
of red supergiants (RSGs) close to the photosphere, to obtain
fundamental parameters, and to locate the stars in the Hertzsprung-Russell (HR) 
diagram. The characterization and location in the HR diagram are 
highly relevant to calibrating stellar evolutionary models for massive stars 
and understanding their further evolution towards Wolf-Rayet (WR) stars or supernovae. 

The fundamental parameters and the atmospheric structure of RSGs have not been characterized well yet. Recent spectrophotometric observations compared to 
MARCS model atmospheres caused a dramatic revision of the location of RSGs 
in the HR diagram (Levesque et al. \cite{Levesque2005}, 
Massey et al. \cite{Massey2006}).
An initial interferometric survey of RSGs of spectral types G8--M2 
(van Belle et al. \cite{Belle2009}) is in broad agreement with these 
new spectroscopic results, with the exception of K type RSGs. 

Interferometric observations of RSGs
at continuum bandpasses have shown that the continuum photosphere
can be well described by a limb-darkened disk 
(Perrin et al. \cite{Perrin2004}; Ohnaka et al. \cite{Ohnaka2011})
that is consistent with MARCS (Haubois et al. \cite{Haubois2009})
and PHOENIX (Wittkowski et al. \cite{Witt2012}) model atmosphere predictions.
A study of the close red supergiant Betelgeuse 
(Ohnaka et al. \cite{Ohnaka2009}, \cite{Ohnaka2011}) 
suggests there are extended CO layers, which cannot be accounted for 
by classical hydrostatic model atmospheres. Other studies suggest the presence 
of extended and asymmetric water vapor layers in RSGs 
(Perrin et al. \cite{Perrin2004}, \cite{Perrin2005}; 
Wittkowski et al. \cite{Witt2012}). The mechanisms that levitate the mass to radii where the wind can be
accelerated by the radiation pressure on dust grains have not yet been
understood for RSGs, but are crucial for the mass-loss
process (e.g., Bladh \& H\"ofner \cite{Bladh2012}). 

Recently, we used the VLTI/AMBER instrument to study the characteristics of the molecular layers 
of Mira variables and the red supergiant VY CMa, and demonstrated that this 
instrument, with its spectro-interferometric capabilities, is well suited 
to revealing both the angular diameter of the continuum photosphere at 
bandpasses with little contamination by molecular bands and the structure 
of the extended atmosphere 
(Wittkowski et al. \cite{Witt2011}, \cite{Witt2012}). 
In this paper we extend VLTI/AMBER studies to a larger sample of RSGs
covering spectral types M1--M5, extending the sample of van Belle et al. (\cite{Belle2009}) to cooler spectral types. 

The remainder of this paper is structured as follows. In Sect. 2 we 
describe our AMBER observations, the data reduction, and the first results.  
In Sect. 3 we explain the atmospheric model used. In Sect. 4 we report 
the results obtained from the model fitting  and estimate the 
fundamental parameters. In Sect. 5 . we discuss the atmospheric structure 
of our targets.
Finally, in Sect. 6 we summarize our conclusions.

\section{Observations, data reduction, and results}

We observed the RSGs AH~Sco (Simbad spectral type M4-5~Iab), UY~Sct (M4~Ia), 
and KW~Sgr (M1.5~Iab) with the ESO Very Large 
Telescope Interferometer (VLTI), utilizing three of the Auxiliary Telescopes 
of 1.8\,m diameter. We used the “Astronomical Multi-BEam combineR” (AMBER) 
with the external fringe tracker FINITO (Petrov et al. \cite{Petrov2007}). 
We worked in medium-resolution mode (R~$\sim$~1500) in the 
K-2.1\,$\mu$m and K-2.3\,$\mu$m bands (together covering wavelengths 
between 1.9\,$\mu$m and 2.5\,$\mu$m). We made our observations as sequences 
of cal-sci-cal (sci is the target and cal the calibrator), observing 
five scans for each of them. The log of the observations is reported 
in Table~\ref{Log_obs}. The integration time (DIT) of each frame was 500\,ms. The used calibrators 
were chosen from the ESO Calibrator Selector CalVin based on the catalog of 
Lafrasse et al. (\cite{Lafrasse2010}). We chose $\tau$~Sgr 
(RA 19 06 56.4, DEC -27 40 13, spectral type K1 III and angular 
diameter 3.83$\pm$0.04\,mas) as an interferometric calibrator for 
AH~Sco and UY~Sct. For KW~Sgr we chose HR~6583 (RA~17~42~51.1, 
DEC~-36~56~44, spectral type K5III and angular diameter 2.48$\pm$0.18 mas) 
and 11~Sgr (RA~18~11~43.3 , DEC~-23~42~04, spectral type K0~III and 
angular diameter 1.43$\pm$0.10\,mas).  

\begin{table}
\caption{VLTI/AMBER observations}
\centering
\begin{tabular}{lcccc}
\hline
\hline
Target  & Date & Mode & Baselines & PA  \\
(Sp. type) &  & K- & BL1/BL2/BL3 & \\
   &  & $\mu$m &  m  & deg         \\
\hline
AH Sco & 2010-06-04 & 2.1  & 40.6/40.4/62.3 & 168/-51/-91  \\
(M4-5) & 2010-06-04 & 2.3  & 40.8/45.2/66.7 & 159/-63/-99\\
UY Sct & 2010-06-04 & 2.1  & 37.6/43.4/58.5 & 163/-64/-103 \\
 (M4)  & 2010-06-04 & 2.3  & 36.9/37.4/49.8 & 172/-56/-102\\
KW Sgr & 2012-04-21 & 2.1  & 62.9/71.5/70.1 & 64/-178/129\\
(M1.5) & 2012-04-21 & 2.3  & 63.7/71.4/71.5 & 73/-170/137\\
\hline
\end{tabular}
\tablefoot{The AMBER instrument mode is K-2.1\,$\mu$m (medium-resolution mode 
with range 1.92-2.26\,$\mu$m) or K-2.3\,$\mu$m (2.12-2.47\,$\mu$m). The baseline 
is the projected baseline length for the AT VLTI baseline used. }
\label{Log_obs}
\end{table}

\begin{figure*}
\centering
\includegraphics[width=0.49\hsize]{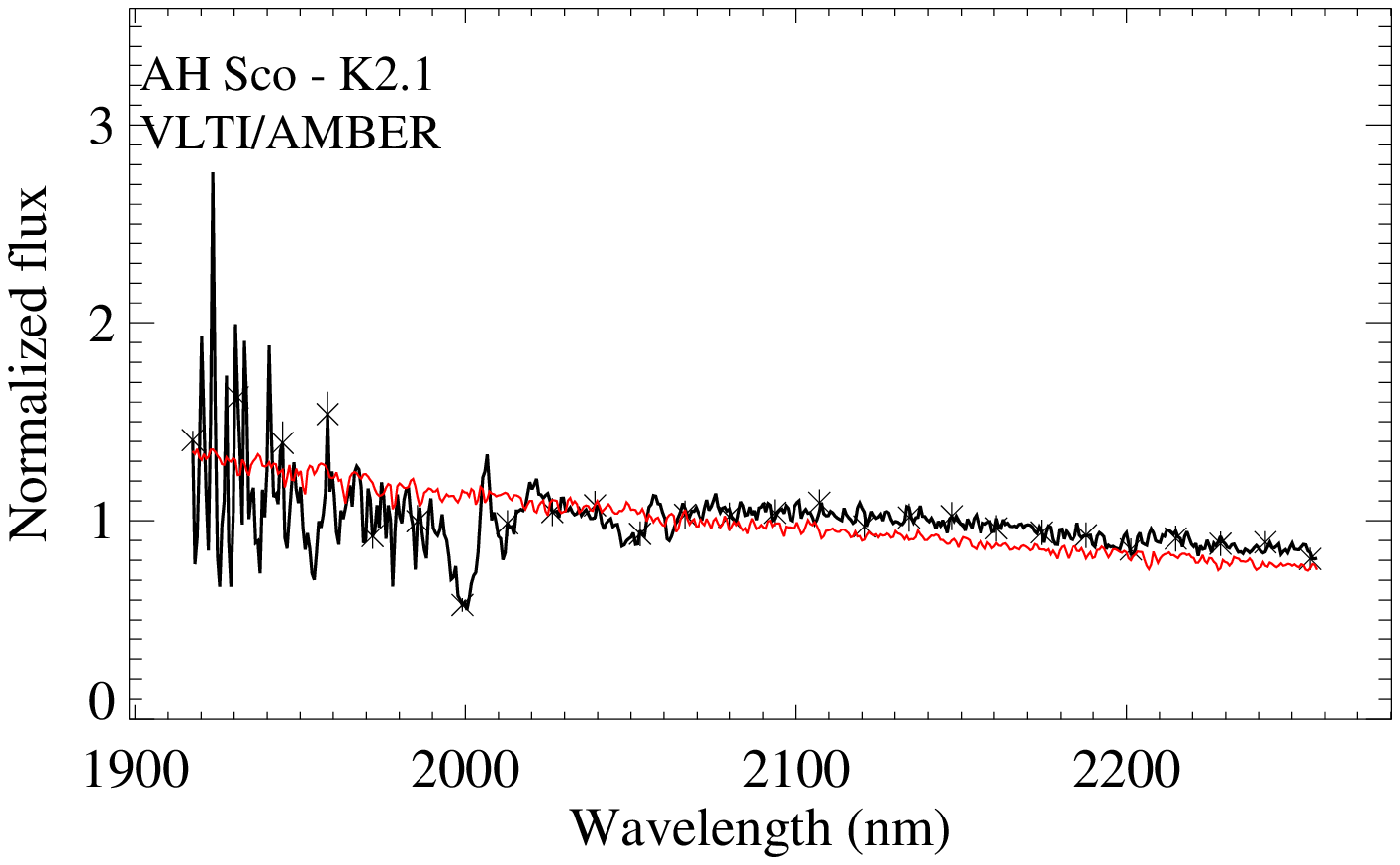}
\includegraphics[width=0.49\hsize]{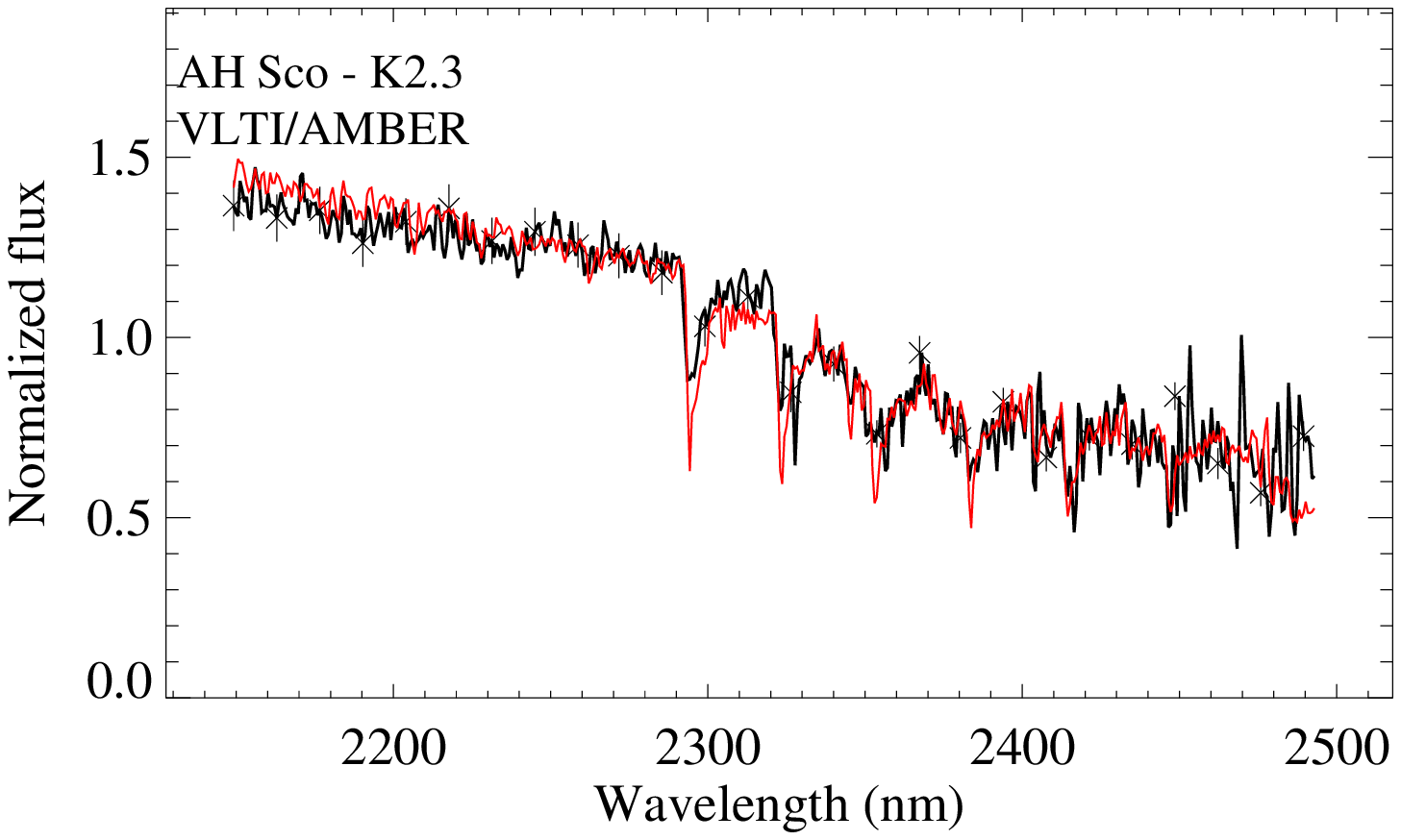}
\includegraphics[width=0.49\hsize]{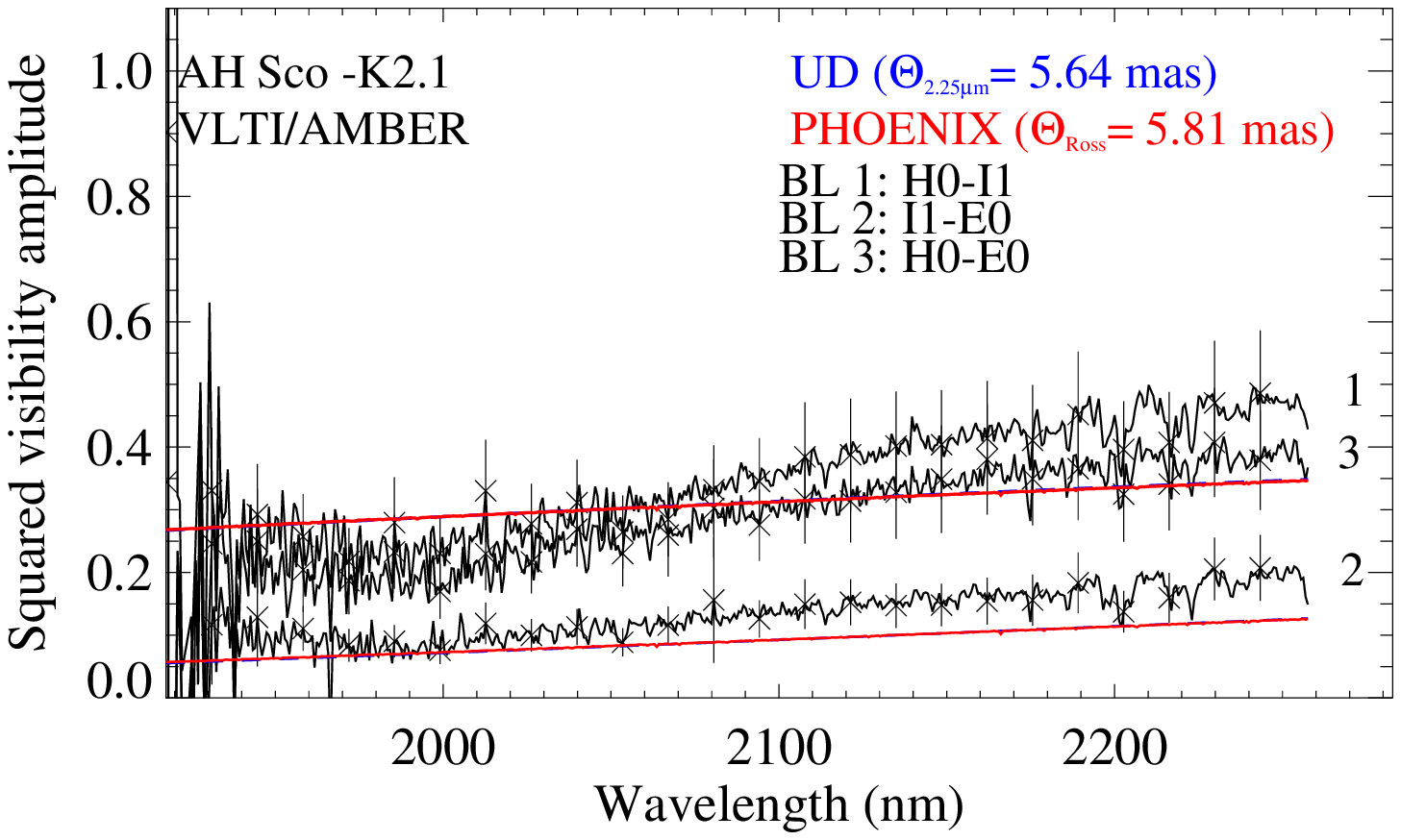}
\includegraphics[width=0.49\hsize]{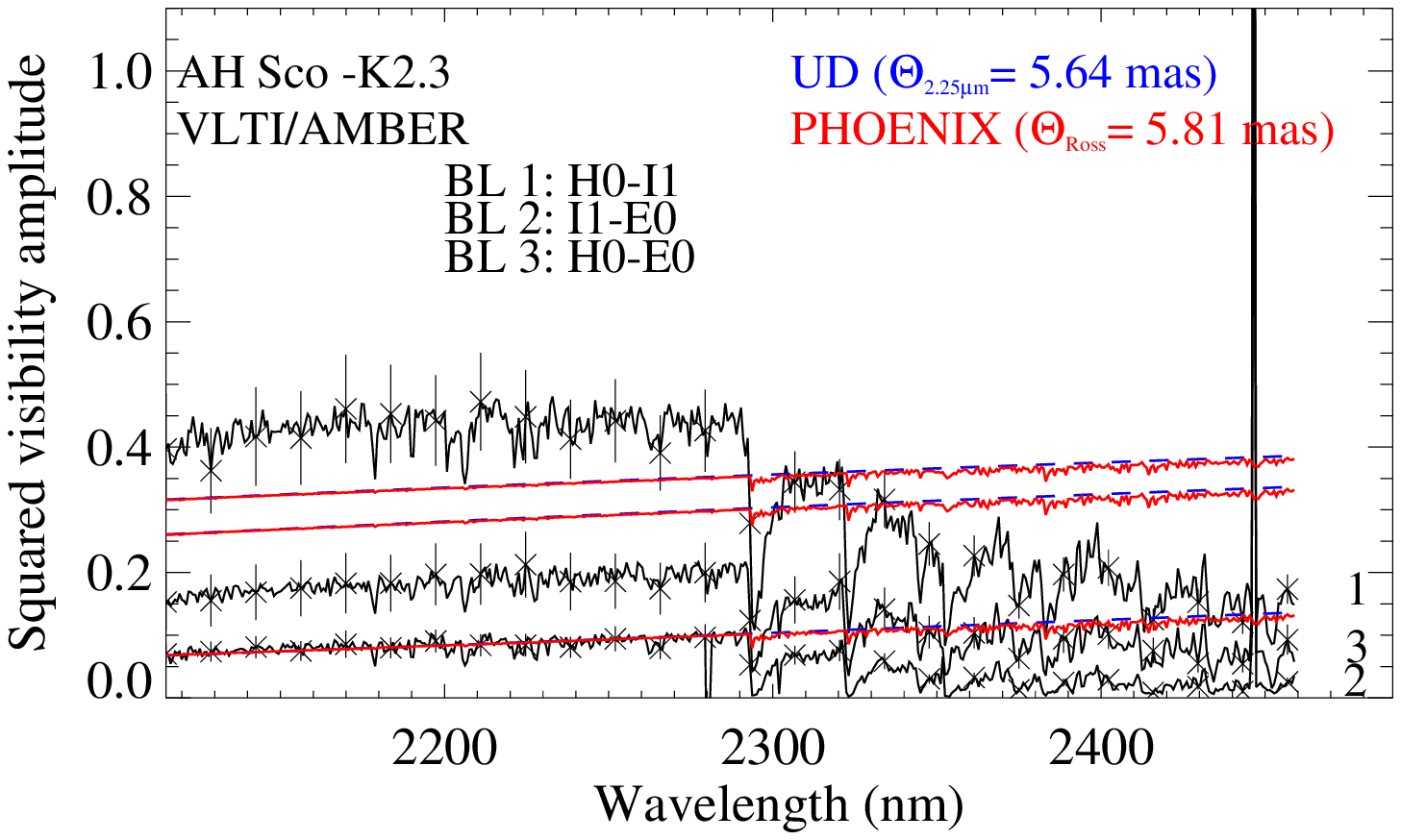}
\includegraphics[width=0.49\hsize]{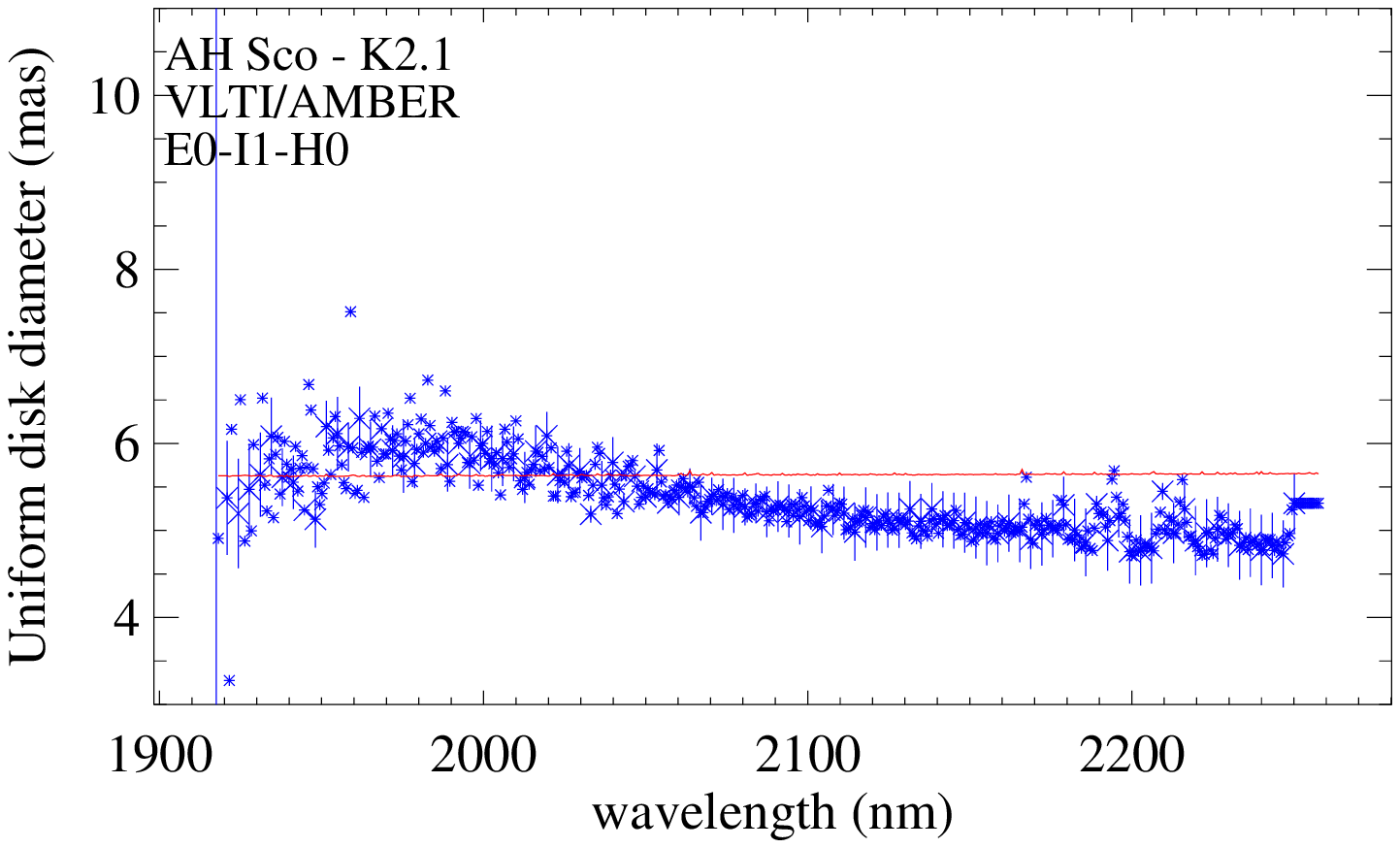}
\includegraphics[width=0.49\hsize]{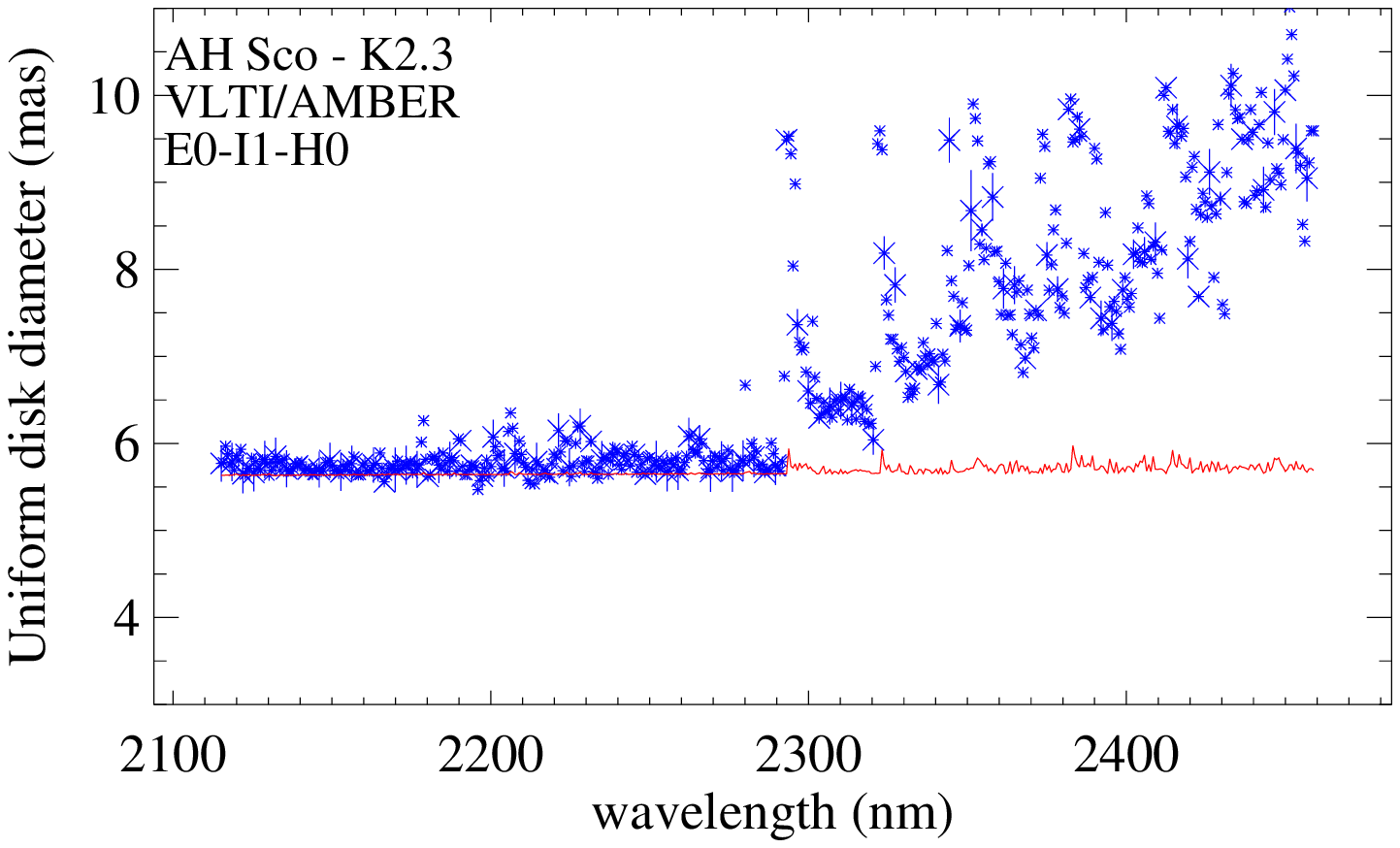}
\includegraphics[width=0.49\hsize]{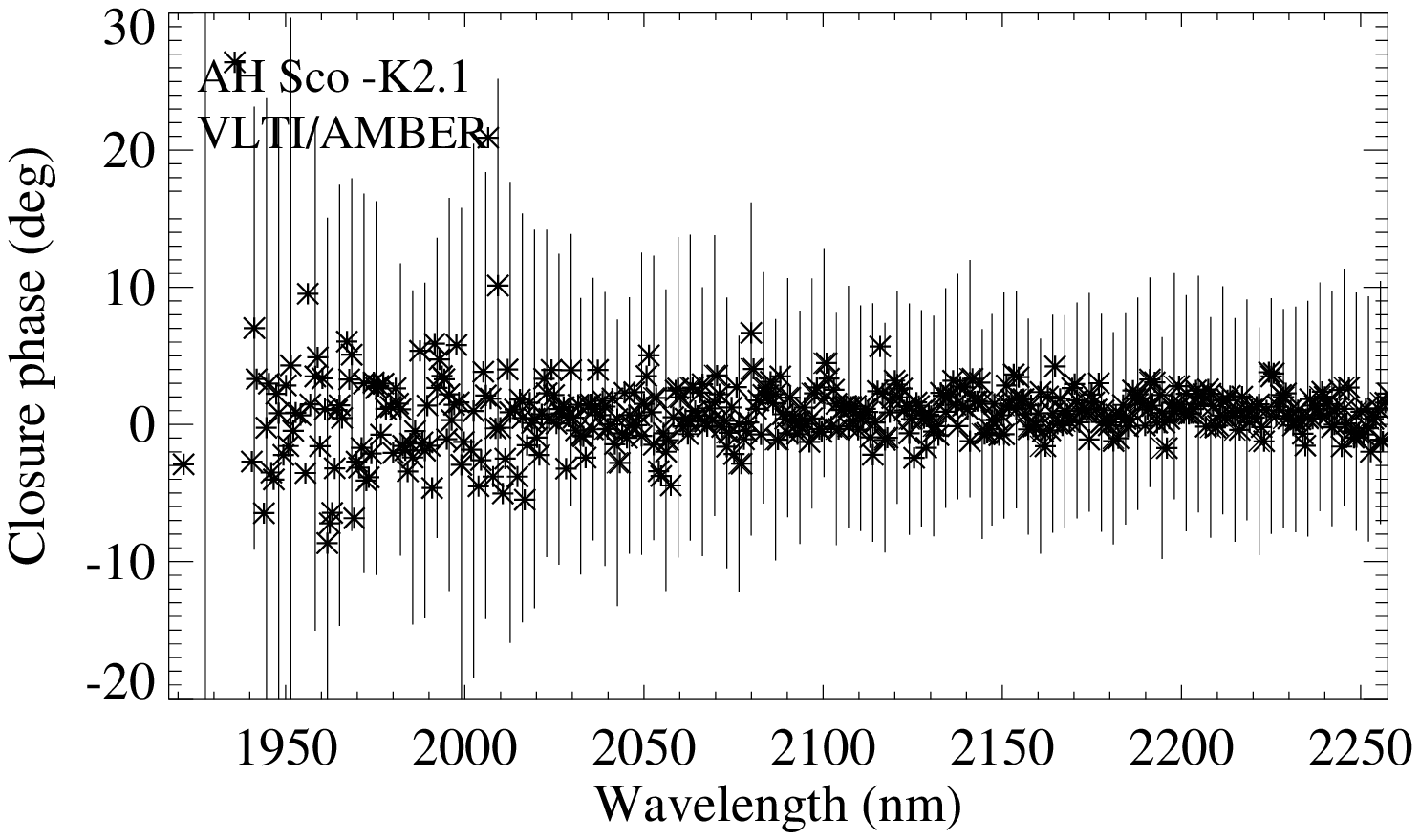}
\includegraphics[width=0.49\hsize]{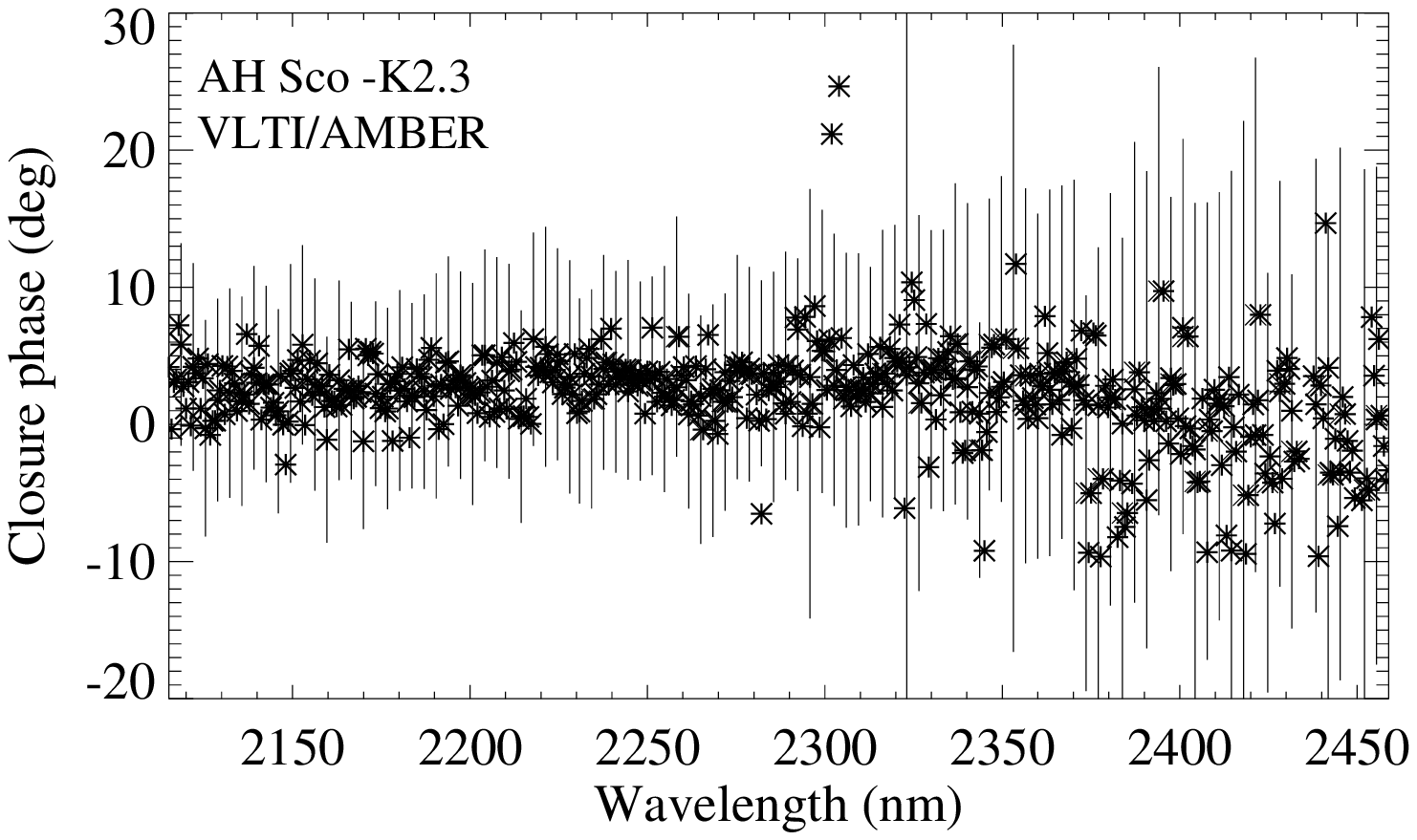}
\caption{Left: Observed (black) normalized flux, 
squared visibility amplitudes,  UD diameters predicted form our data (blue) and from 
the best-fit PHOENIX model (red), and closure phases
in degree (from top to bottom) of AH~Sco obtained with the 
MR-K 2.1\,$\mu$m setting. Right: Same as left but 
obtained with the MR-K 2.3\,$\mu$m setting. 
The blue curves show the best-fit UD model, and the red curves
the best-fit PHOENIX model prediction. For the baseline lengths and position angles, see Table \ref{Log_obs}. In the 2nd panel on the left, two model lines overlap, corresponding to baselines of 40.6m and 40.4m.
}
\label{resul_AHSco_fit}
\end{figure*} 

\begin{figure*}
\centering
\includegraphics[width=0.49\hsize]{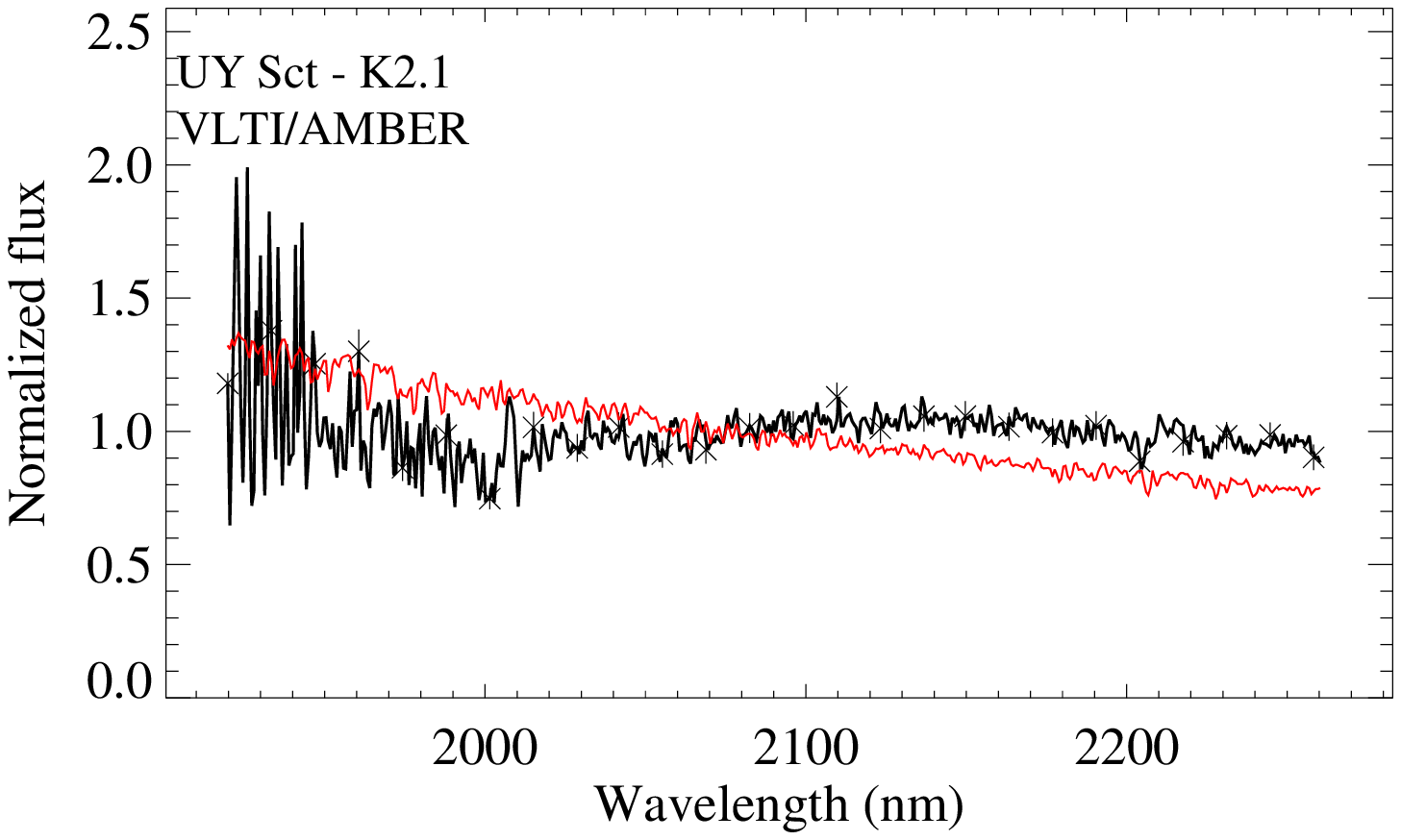}
\includegraphics[width=0.49\hsize]{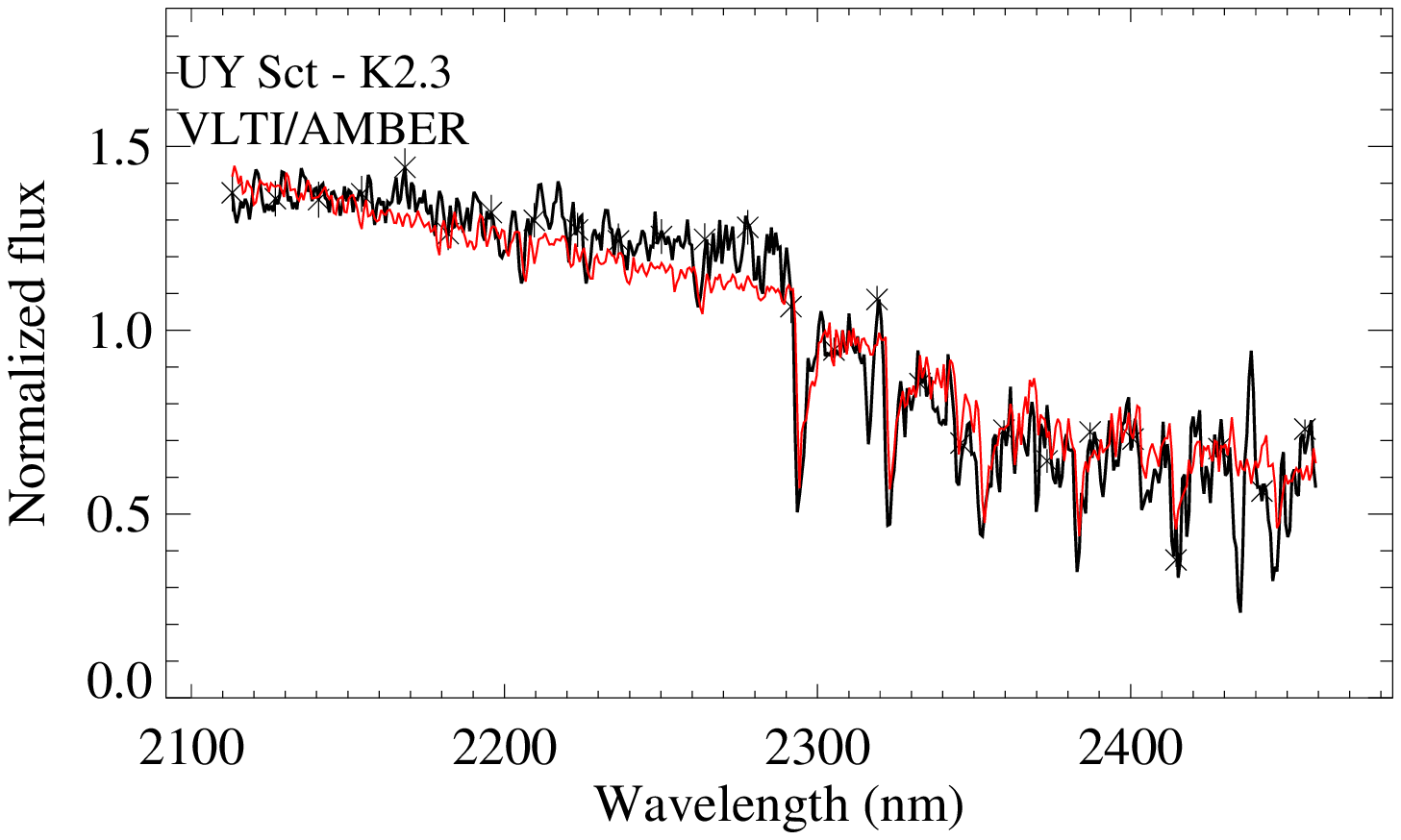}
\includegraphics[width=0.49\hsize]{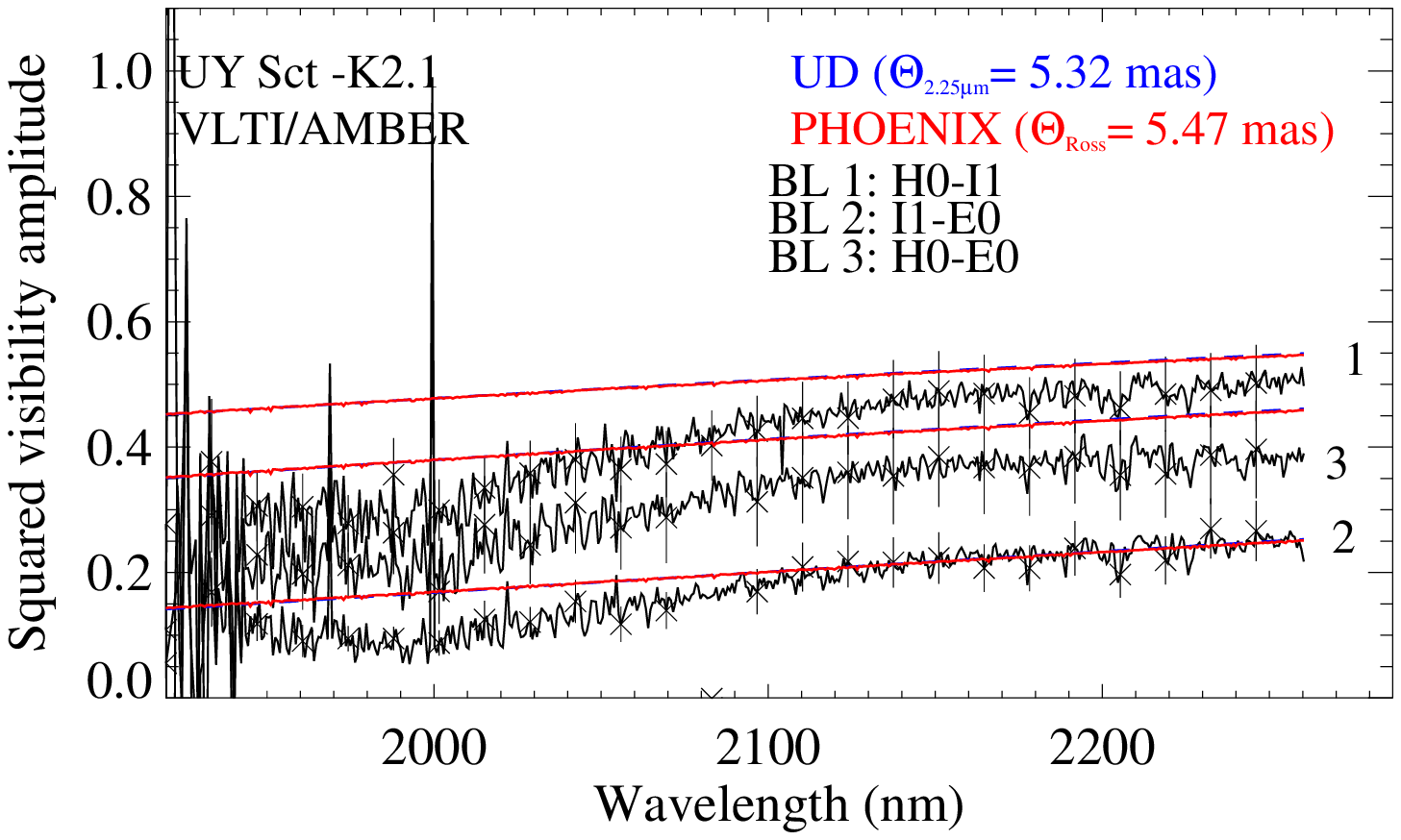}
\includegraphics[width=0.49\hsize]{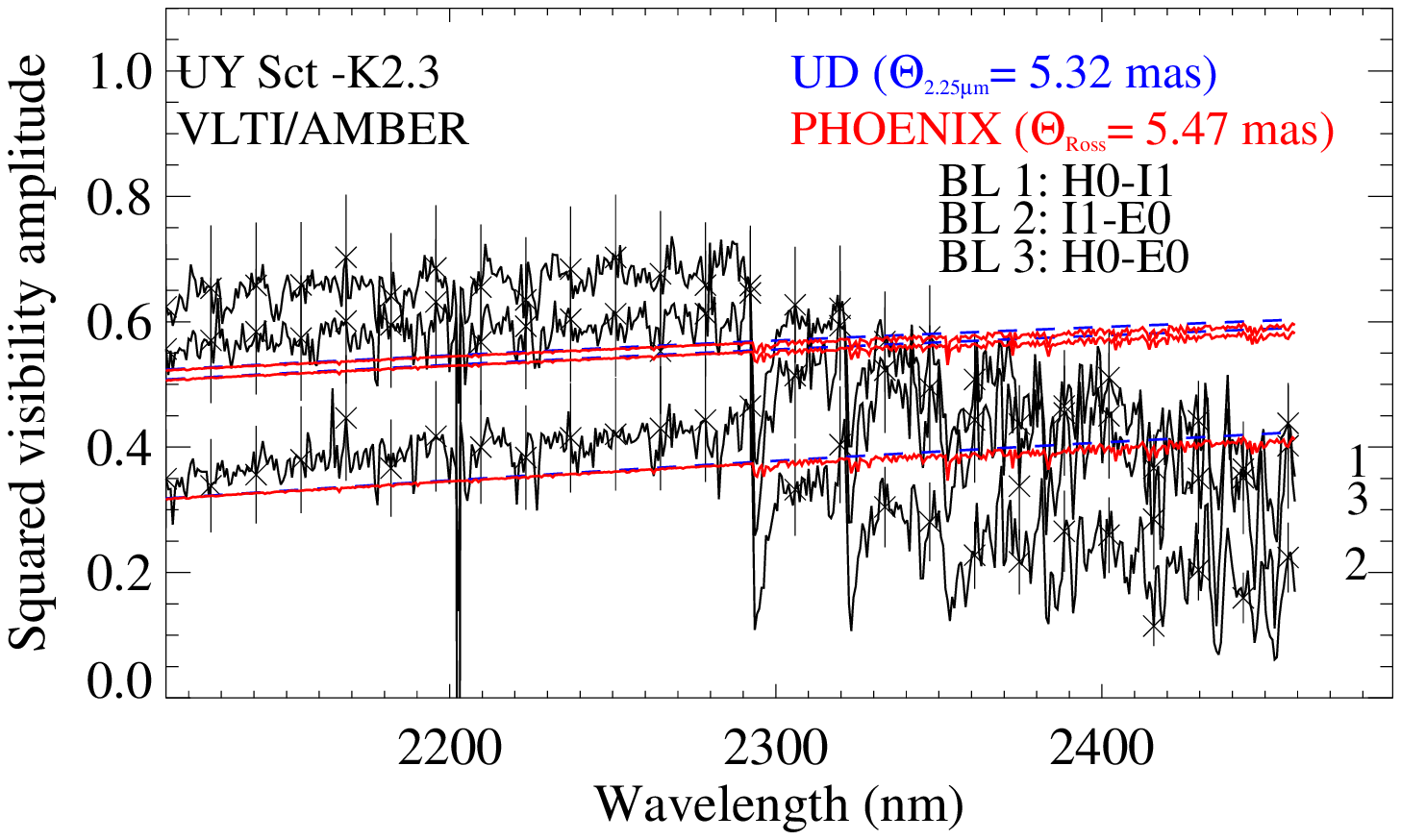}
\includegraphics[width=0.49\hsize]{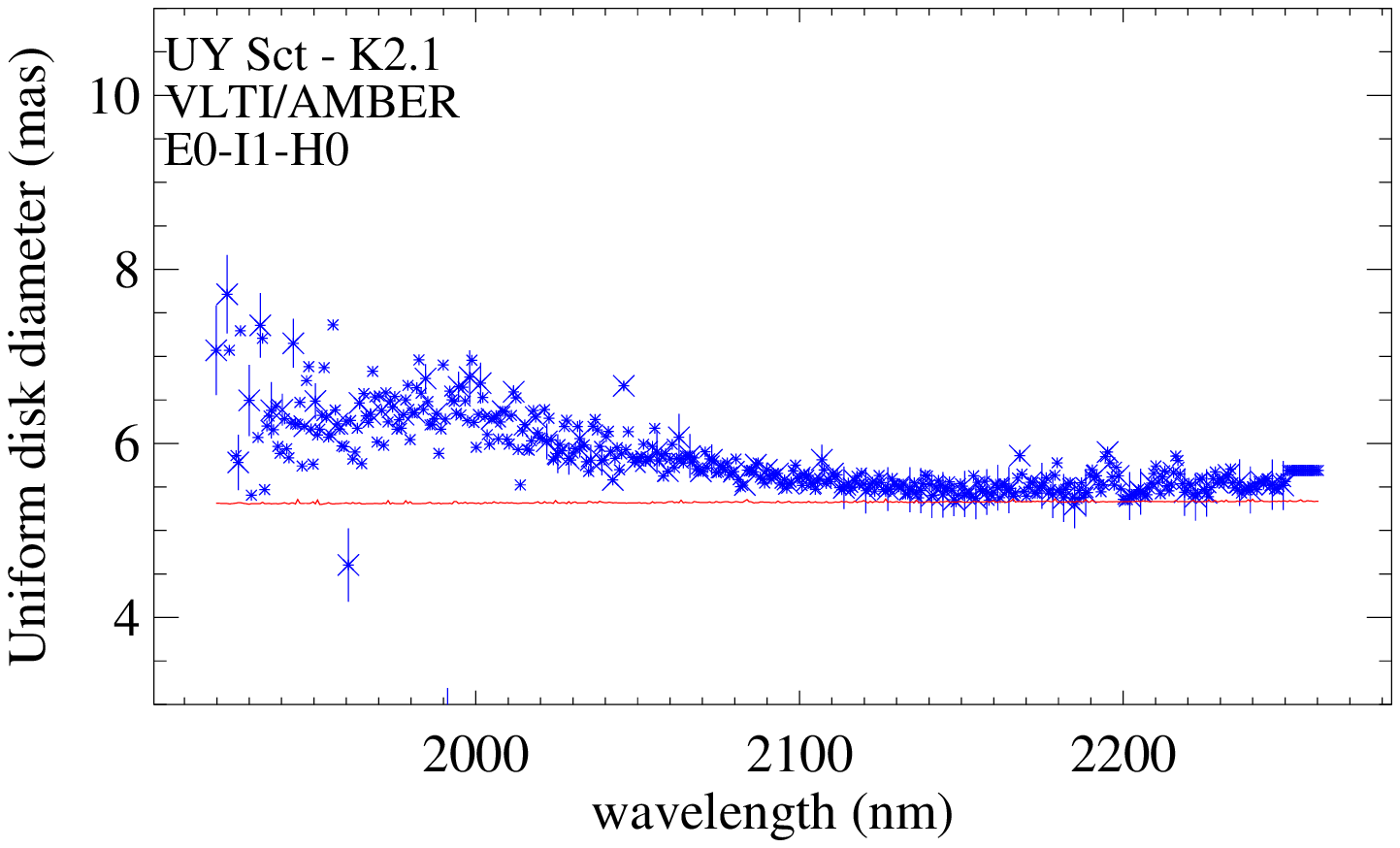}
\includegraphics[width=0.49\hsize]{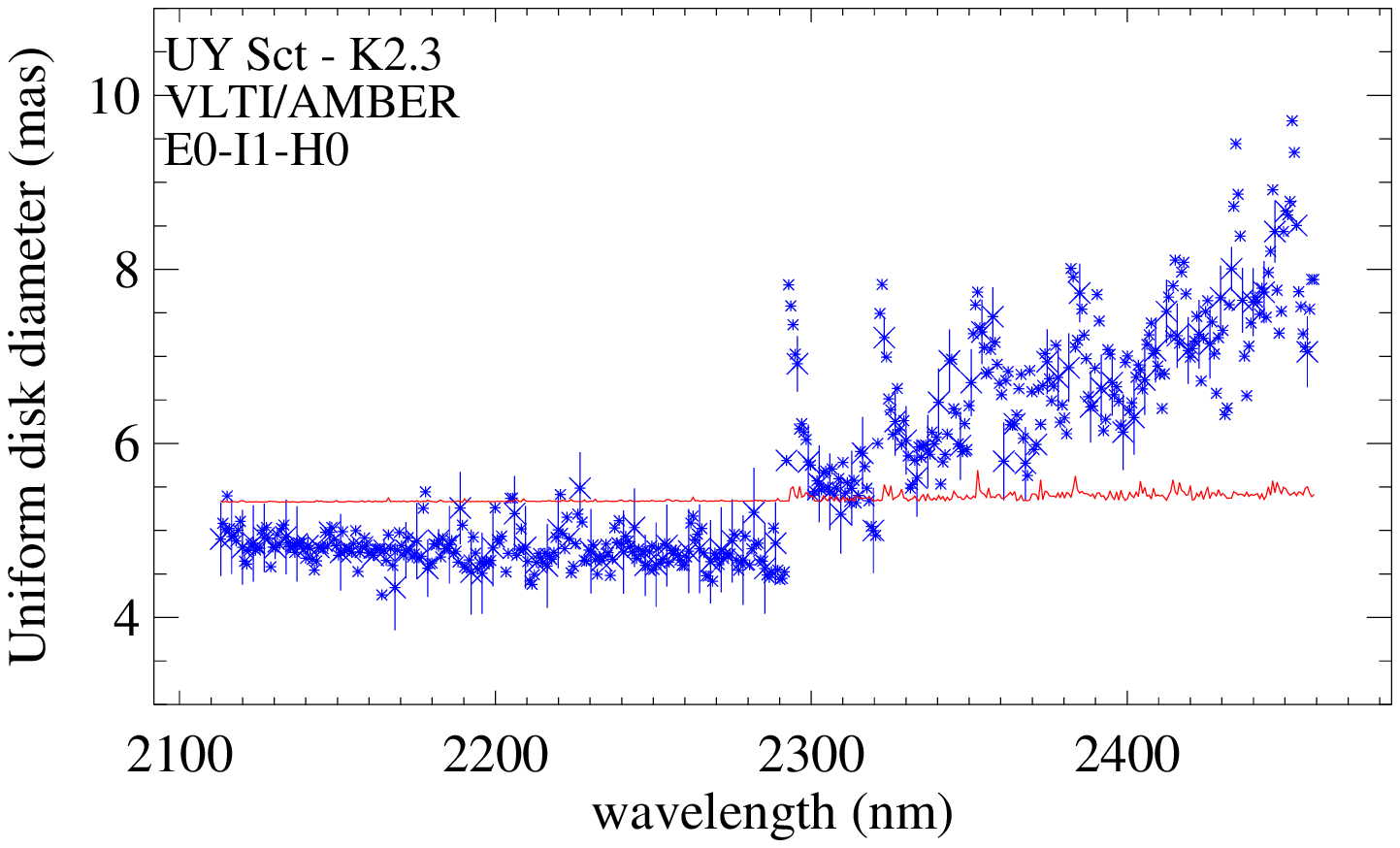}
\includegraphics[width=0.49\hsize]{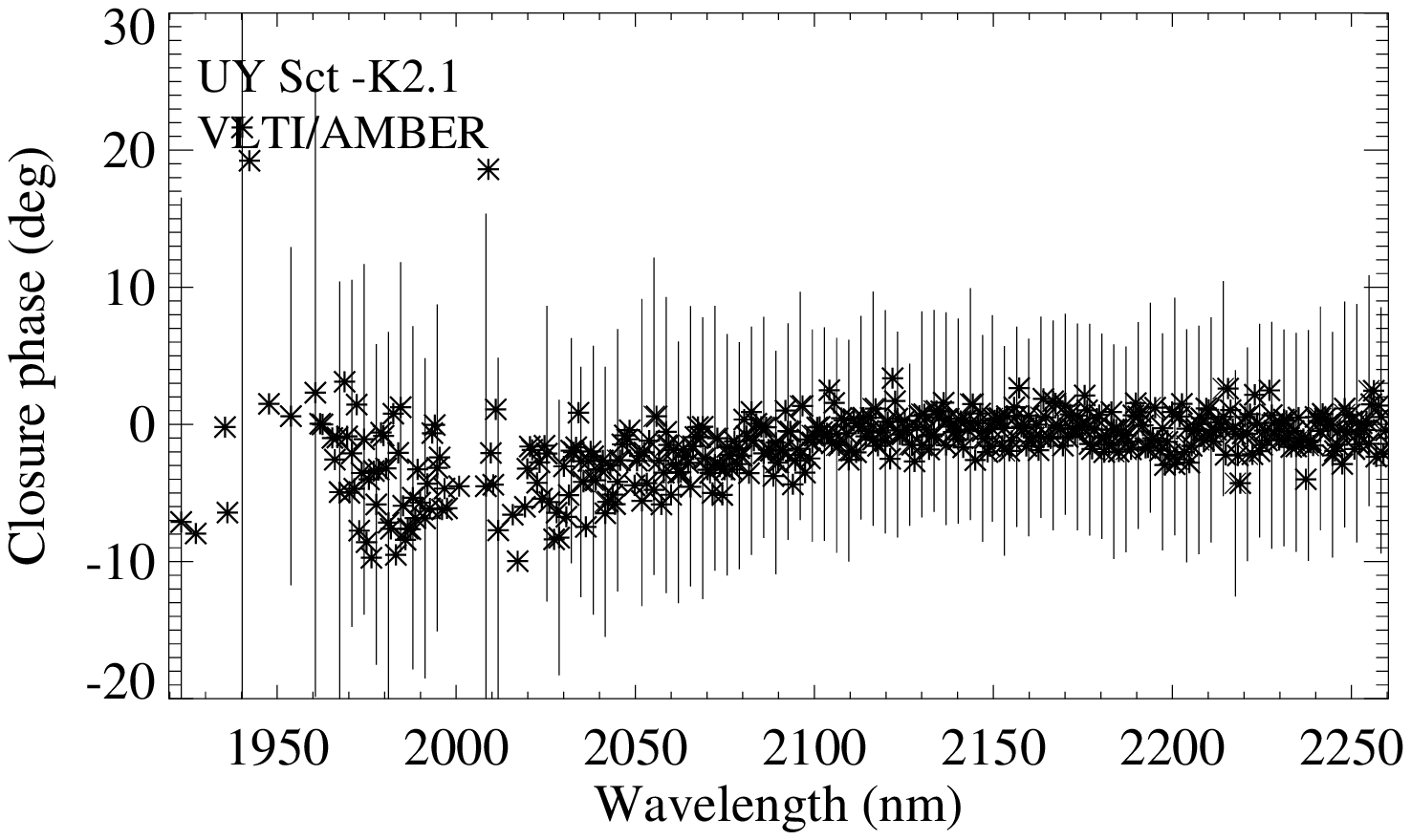}
\includegraphics[width=0.49\hsize]{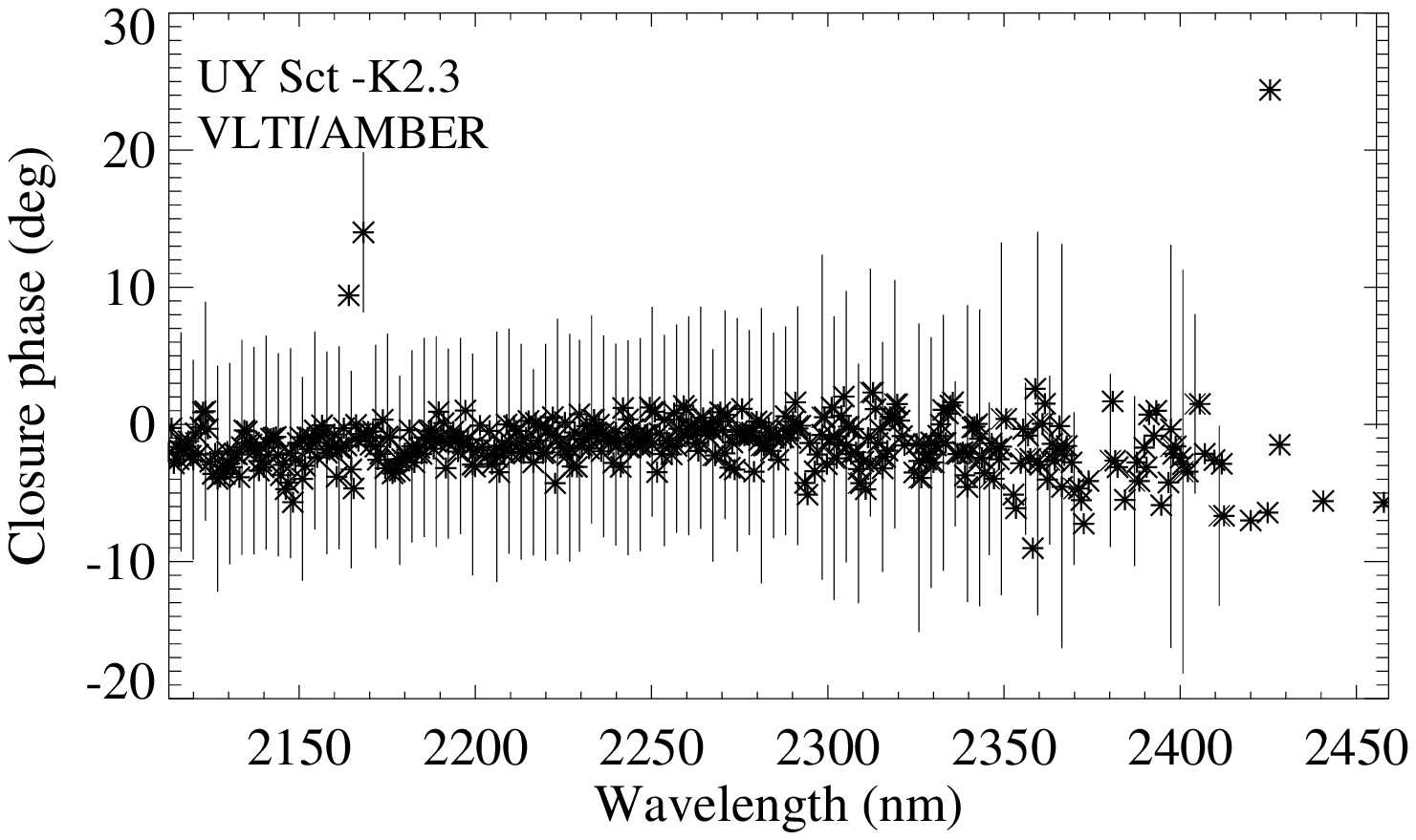}
\caption{
As Fig.~\protect\ref{resul_AHSco_fit}, but for UY~Sct.}
\label{resul_UYSct_fit}
\end{figure*} 

\begin{figure*}
\centering
\includegraphics[width=0.49\hsize]{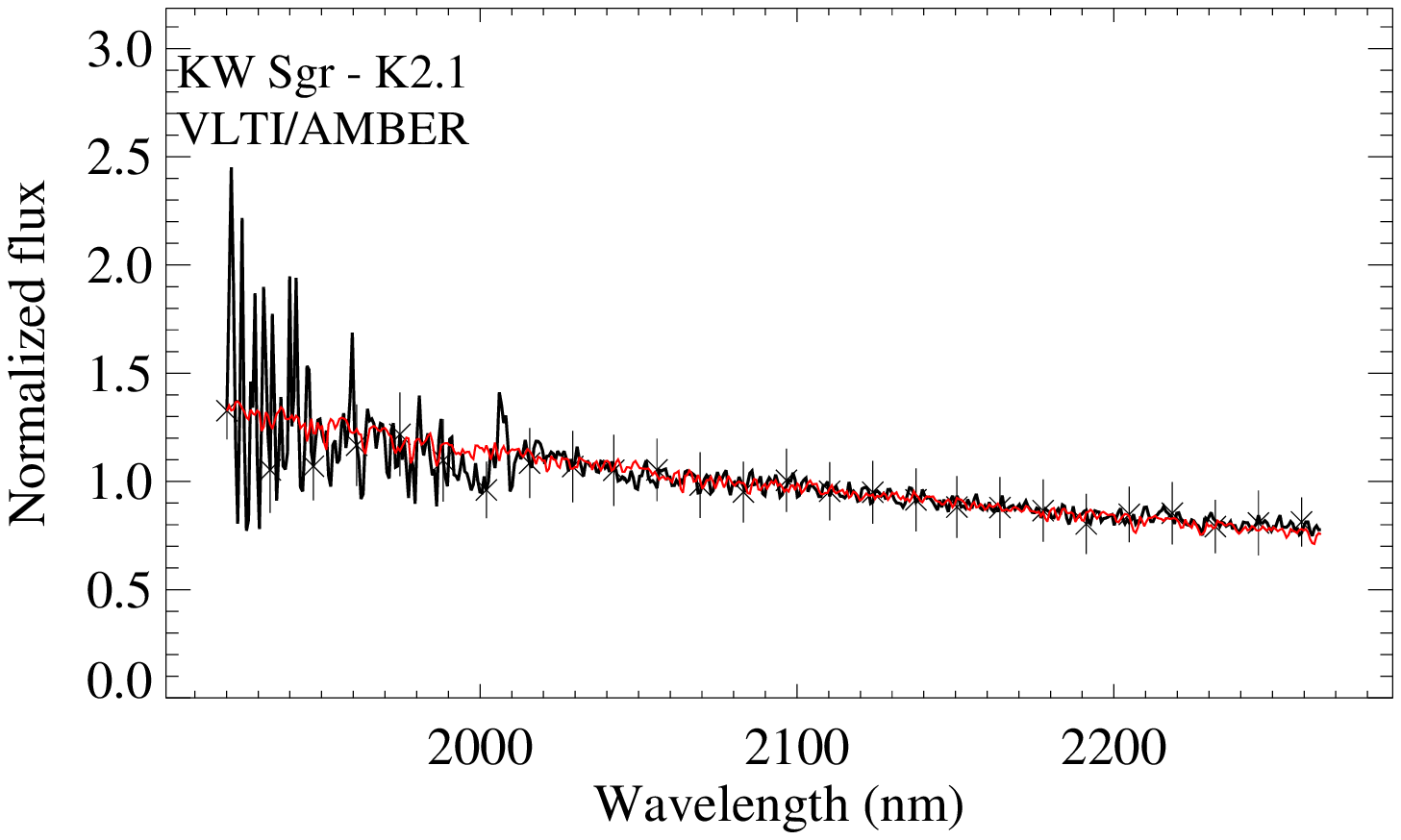}
\includegraphics[width=0.49\hsize]{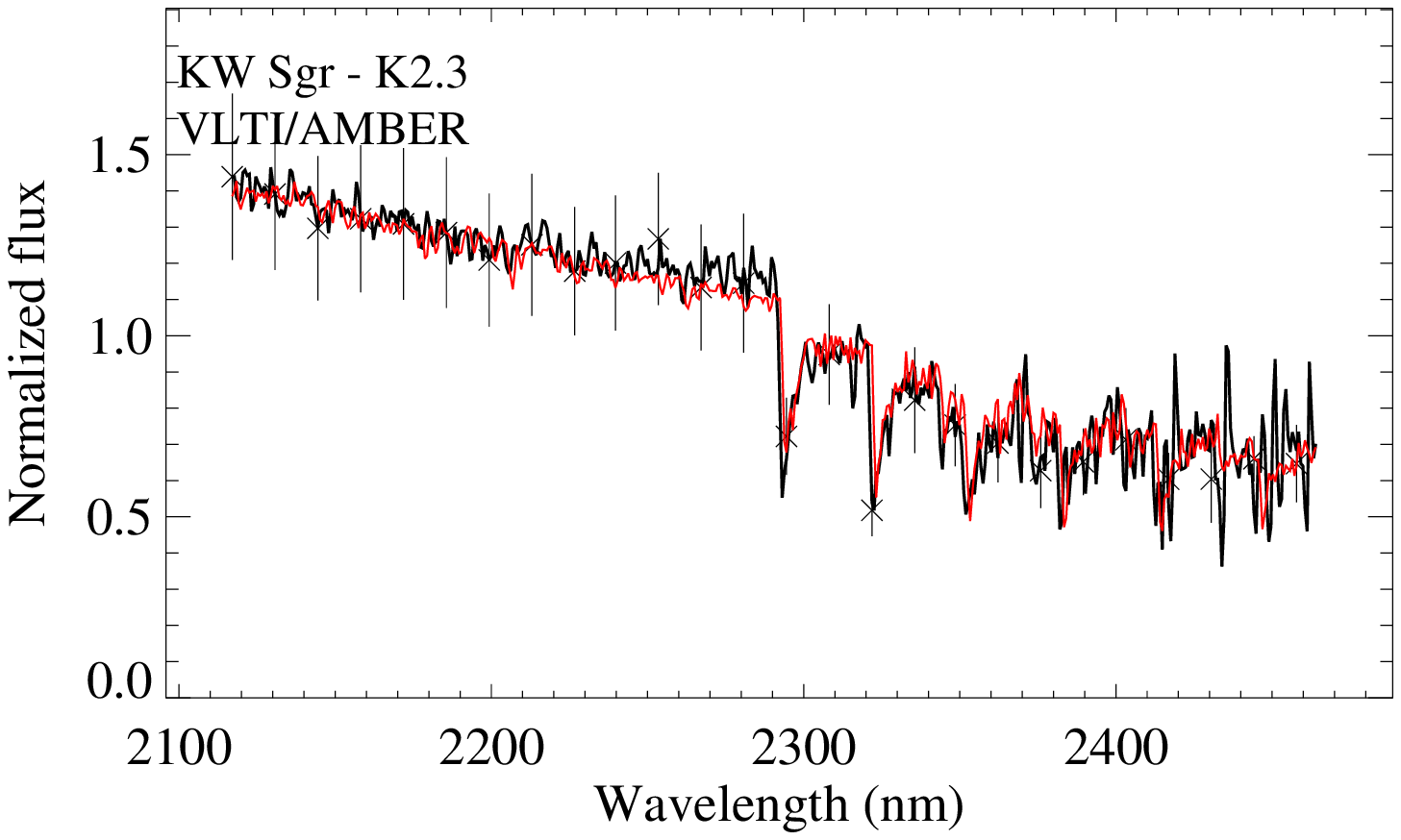}
\includegraphics[width=0.49\hsize]{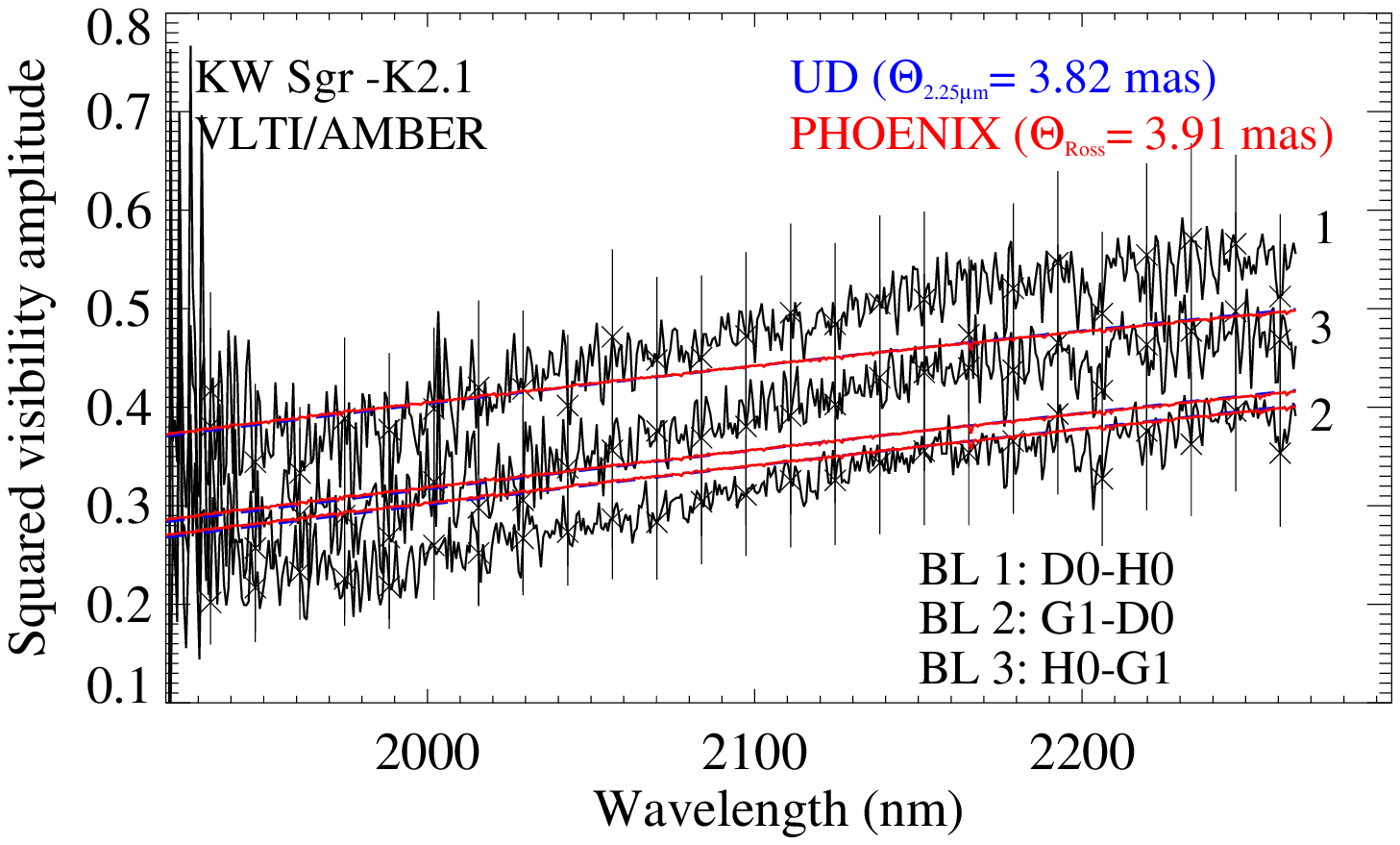}
\includegraphics[width=0.49\hsize]{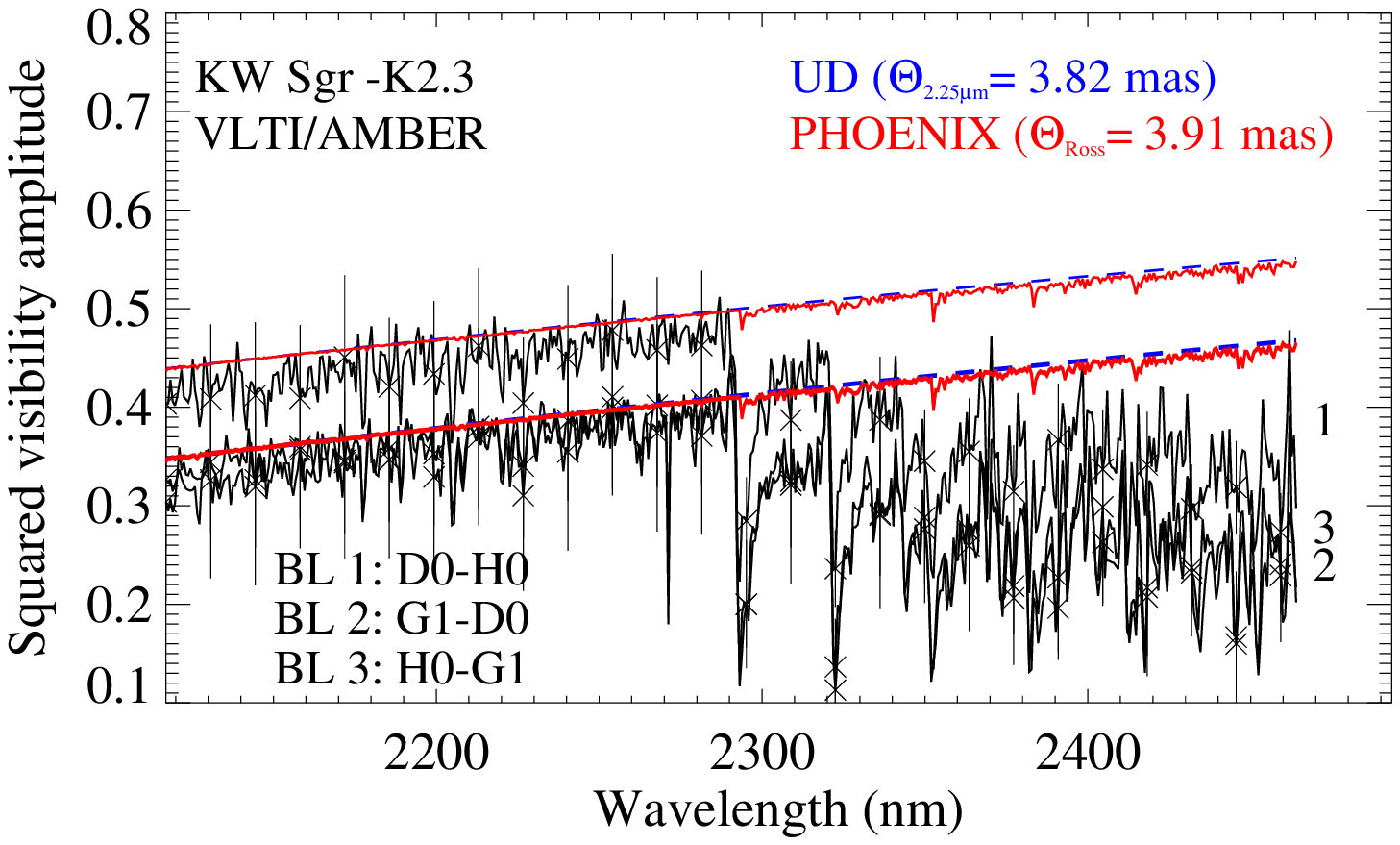}
\includegraphics[width=0.49\hsize]{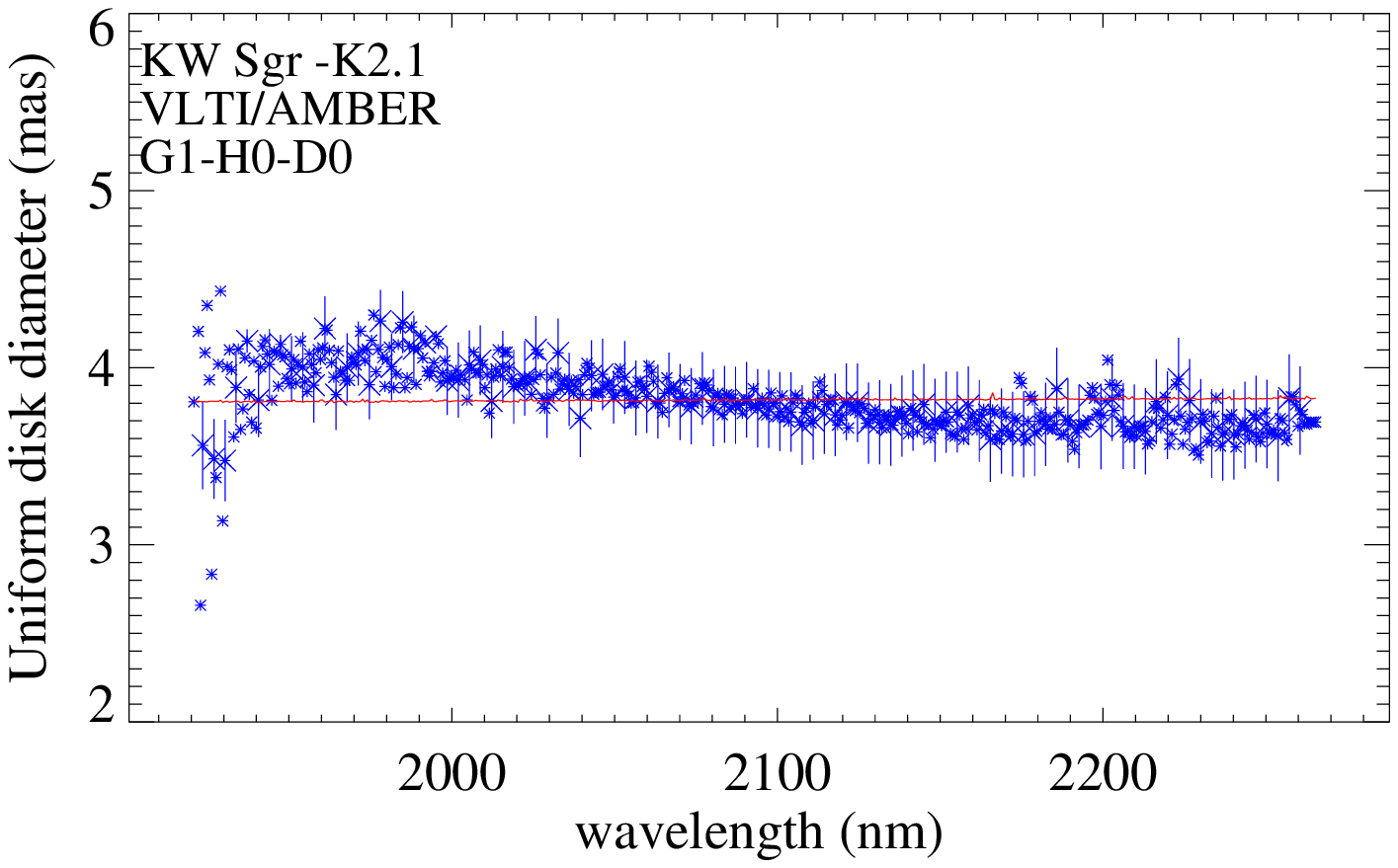}
\includegraphics[width=0.49\hsize]{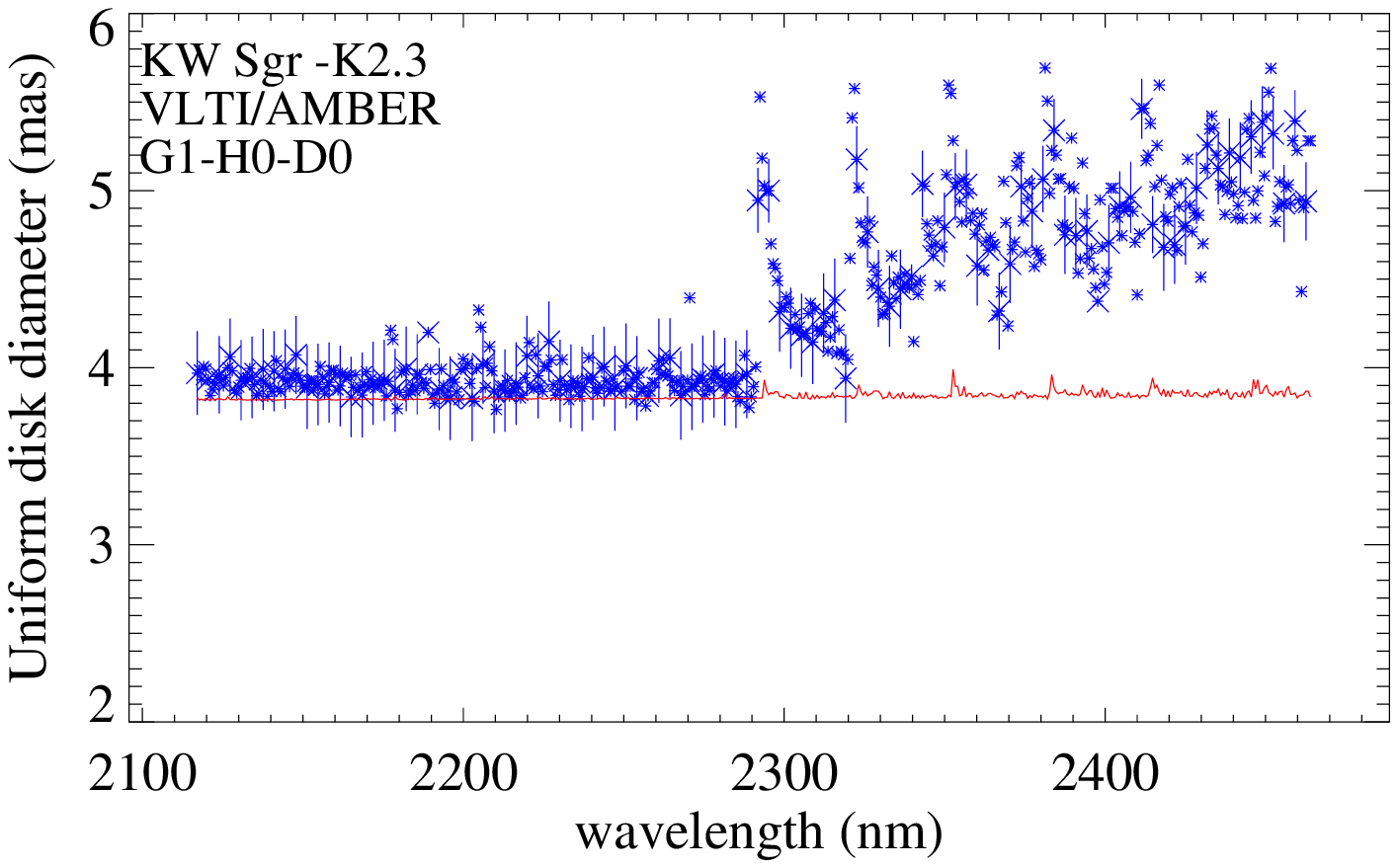}
\includegraphics[width=0.49\hsize]{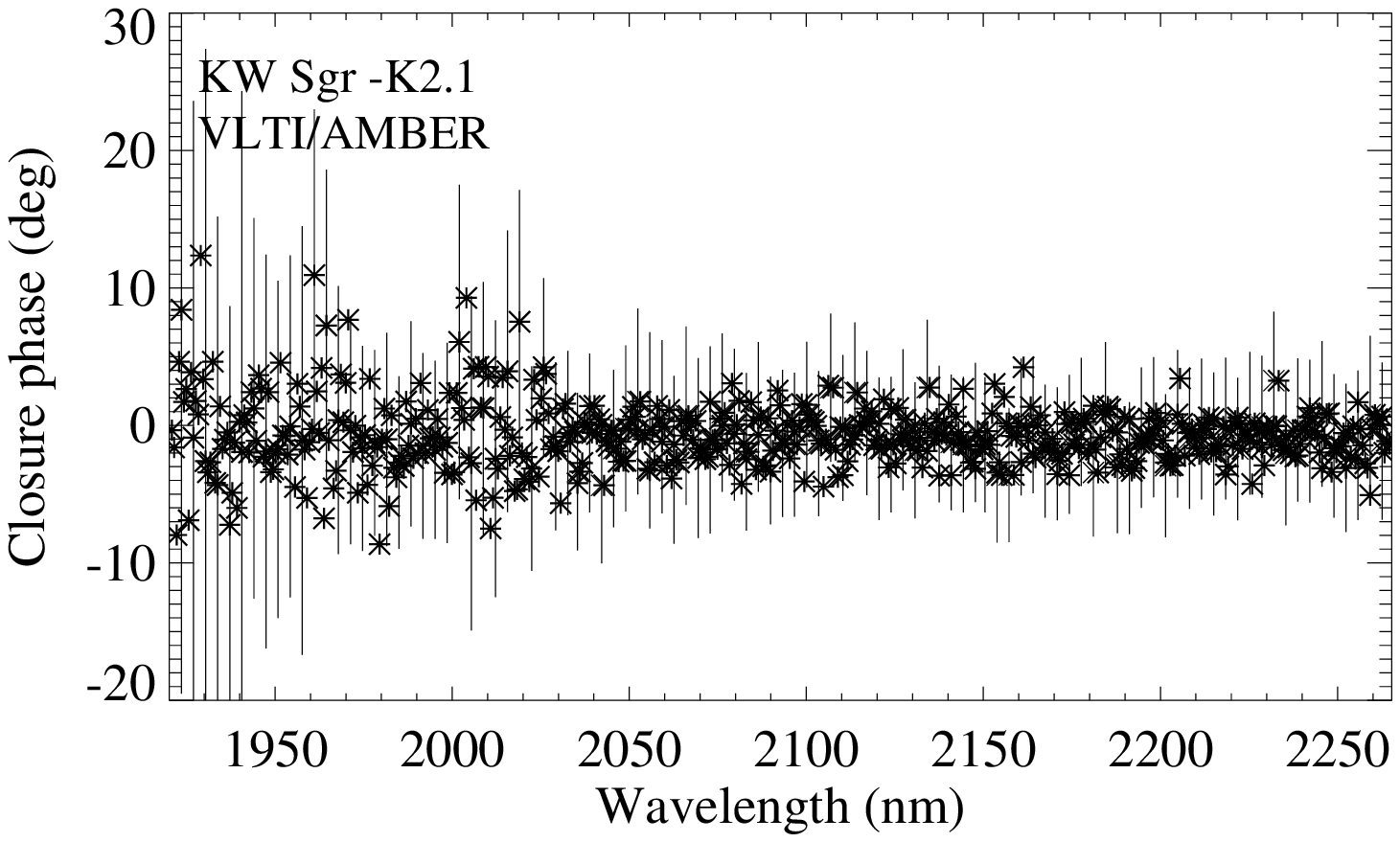}
\includegraphics[width=0.49\hsize]{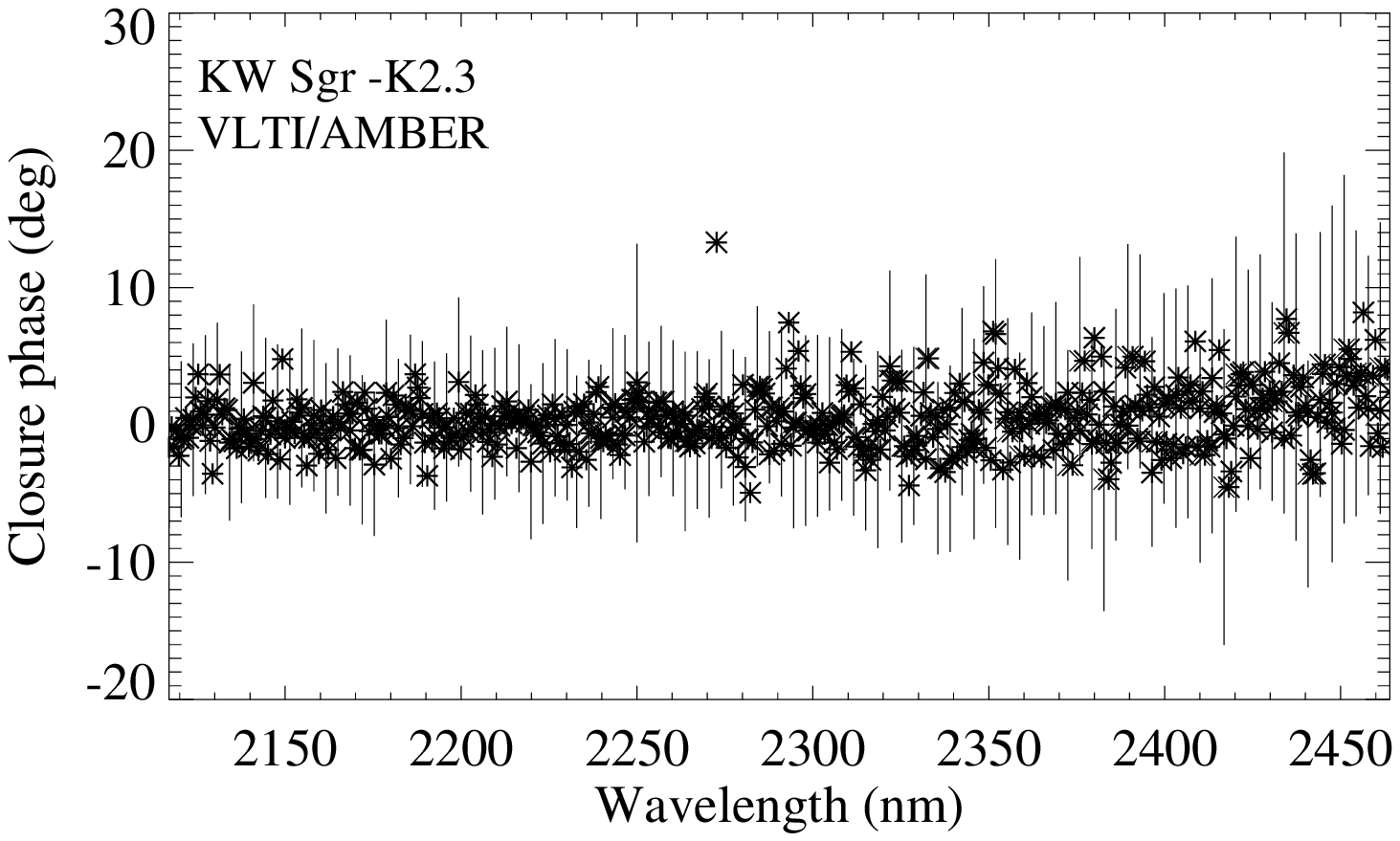}
\caption{
As Fig.~\protect\ref{resul_AHSco_fit}, but for KW~Sgr.}
\label{resul_KWSgr_fit}
\end{figure*} 

Raw visibility and closure phase values were obtained from our AMBER 
data using the latest version of the amdlib data reduction package 
(Tatulli et al. \cite{Tatulli2007}; Chelli et al. \cite{Chelli2009}). 
We appended all scans taken consecutively from the same source.  
Later, we selected and averaged the resulting visibilities of each frame 
using different criteria. In our case, the criteria were based on the baseline flux and on
the fringe signal-to-noise ratio (SNR, for more information 
see the AMBER Data Reduction Software User Manual\footnote{http://www.jmmc.fr/doc/approved/JMMC-MAN-2720-0001.pdf}). The baseline flux of baseline i-j
is defined as $\sqrt{f_i f_j}$, with $f_i$ and $f_j$ being the flux from
stations i and j comprising the baseline i-j. We selected all
frames that have a baseline flux that is, at least, three times larger than the associated
noise and 80\% of the remaining frames with best fringe SNR. After that, we performed 
the absolute wavelength calibration by correlating the AMBER flux spectra 
with a reference spectrum that included the AMBER transmission curves, 
the telluric spectrum estimated with ATRAN (Lord \cite{Lord1992}), and the 
expected stellar spectrum, using the BS~4432 spectrum from 
Lan\c con \& Wood (\cite{Lancon2000}). We selected this star because 
it has a spectral type similar to our calibrators (the spectral type of our 
calibrator are: K1 III, K5 III, and K0 III and the spectral type of 
BS 4432 is K4.5 III). Then, we calibrated the flux and the 
visibility spectra. A relative flux calibration of the target was performed 
by using the calibrator and the BS~4432 spectrum.  
Calibrated visibility spectra were obtained by using the average of 
two transfer function measurements taken before and after each science 
target observation. For the observation of AH Sco with the 
K-2.3 setting, the calibrator scans taken before had a low quality. In this 
case we used only one calibrator. The error of the transfer function ($\Delta T$) 
was calculated as a function of two terms (T$_{A}$ and T$_{B}$). The first one 
was obtained by error propagation ($T_{A}=1/V_{inst}*ΔV^{m}_{cal}$), and 
the second one as $T_{B}=|T_{1}-T_{2}|/2$, where T$_{1}$ and T$_{2}$ are the 
transfer functions of cal$_{1}$ and cal$_{2}$. When we used only one calibrator, 
we set T$_{B}$ to a value of 0.05, as obtained from our other observations. 

Figures~\ref{resul_AHSco_fit}, \ref{resul_UYSct_fit}, and 
\ref{resul_KWSgr_fit} show the resulting flux and visibility data 
of AH~Sco, UY~Sct, and KW~Sgr, respectively,
together with predictions by model atmospheres, which will be discussed 
in the following sections.

The normalized flux spectra (top panels of 
Figs.~\ref{resul_AHSco_fit}-\ref{resul_KWSgr_fit}) show typical 
spectra of red supergiants, as observed for instance by 
Lan\c con et al. (\cite{Lancon2007}). They show a decreasing flux 
between 1.9\,$\mu$m and 2.5\,$\mu$m and exhibit strong absorption 
features of CO. The flux variations at wavelengths below about 2.05\,$\mu$m
are due to a higher noise level, possibly caused by the lower
atmospheric transmission. These features are not seen in the visibility. The visibility curves (2nd panels) show a maximum near 2.25\,$\mu$m, and a decrease towards the water bands 
(centered at 1.9\,$\mu$m) and at the positions of the CO bandheads 
(2.3\,$\mu$m to 2.5\,$\mu$m).

The uniform disk diameter calculated from our data as a function 
of wavelength (3rd panels), exhibits a minimum with a constant diameter
at 2.20-2.25\,$\mu$m, and it increases in the H$_{2}$O band 
and the CO bandheads. In the water band, the increase is about 25\% for AH~Sco, 
16\% for UY~Sct, and  10\% for KW~Sgr compared to the near-continuum bandpass (2.15-2.25\,$\mu$m).
In the CO bandheads it increases by about 65\% for AH~Sco, 
45\% for UY~Sct, and  25\% for KW~Sgr.
The decrease in the visibility and the corresponding increase in the
UD diameter indicates a larger contribution from 
extended intensity at the molecular bands (water vapor and CO). 
These results prove there are extended molecular layers of H$_2$O and 
CO in AH~Sco, UY~Sct, and KW~Sgr. The presence of molecular layers 
in the extended atmospheres of red supergiants is consistent with other 
observations of RSGs, such as VY~CMa, $\mu$~Cep, and Betelgeuse 
(Perrin et al. \cite{Perrin2004}, \cite{Perrin2005};
Tsuji \cite{Tsuji2006}; Ohnaka et al. \cite{Ohnaka2009}, \cite{Ohnaka2011}; Wittkowski et al. \cite{Witt2012}). 

In Figs.~\ref{resul_AHSco_fit}-\ref{resul_KWSgr_fit} (3rd panels), we also 
show the uniform disk diameter predicted by the best-fit PHOENIX models. According to these 
results, the PHOENIX model is significantly more compact than observed. 
In particular, the observed extended water bands are not seen in the PHOENIX model predictions, and the 
observed CO layers are included in the models but significantly more compact compared to our 
observations. These findings are consistent with the observations of AGB stars by 
Marti Vidal et al. (\cite{Marti2011}).

The closure phases (bottom panels) show low values of~$\leq$~10~deg in 
both bands 
(K-2.1\,$\mu$m and K-2.3\,$\mu$m) and do not indicate deviations
from point symmetry. However, because of the limited angular resolution 
of our observations that correspond to visibility measurements within
the first lobe, we cannot exclude asymmetries on scales smaller
than the stellar disk as observed, for instance, by Wittkowski et al.
(\cite{Witt2012}) for VY~CMa.


\section{A new grid of PHOENIX model atmospheres}

After the data reduction, we compared our results to PHOENIX model
atmosphere predictions
to determine the angular diameter of our sources and their fundamental 
parameters. Here, we describe the general properties of our new grid
of PHOENIX model
atmospheres, while the comparison to our data is presented below in 
Sects. 4 and 5. We used version 16.03  of the PHOENIX code (for a general 
description see Hauschildt \& Baron \cite{Hausch1999}). 
The code assumes a hydrostatic atmosphere and local thermodynamic 
equilibrium (LTE) and computes $7\cdot 10^{5}$ atomic lines and 
$9\cdot 10^{7}$ molecular lines.
Comparing results obtained with a spherical geometry and a 
plane-parallel geometry, the difference between the two geometries is 
greater for lower T$_\mathrm{eff}$ (Hauschildt et al. \cite{Hausch1999-2}). 
As our targets are expected to show relatively low effective temperatures
between 3200\,K and 3700\,K, we used a spherical geometry. 

We computed a new grid of a total of 127 PHOENIX model atmospheres with 
effective temperatures ranging between 2500\,K and 3900\,K, 
surface gravities between
log(g)=-0.5 and log(g)=4.0 in steps of 0.5, and masses of 1\,M$_\odot$ and 
20\,M$_\odot$. We also computed models with a low mass because we plan to fit these models to red giants in a future work. We used solar metallicity and a micro-turbulent velocity of 2 km/s for all models.
We tabulated model intensity profiles at 64 viewing angles for wavelengths 
from 1.8\,$\mu$m to 2.5\,$\mu$m in steps of 0.01\,$\AA$. Afterward,
we averaged the monochromatic intensity profiles to match the 
spectral channels of the individual observations
and computed the flux integrated over the stellar disk, as well
as the visibility values for the baseline used (for more information on 
this procedure, see Wittkowski et al. \cite{Witt2003}).

The modeling of atmospheres of cold giant stars, such as AH~Sco, UY~Sct, 
and KW~Sgr, is complicated because of two effects, the treatment of 
molecular opacities and their spherical extension of the atmospheres 
(Hauschildt et al. \cite{Hausch1999-2}). In the next section we discuss 
the comparison to our data and the impact of these effects.


\section{Fundamental parameters}

\begin{table*}
\caption{Photometry datasheet}
\centering
\begin{tabular}{lccccccccccccc}
\hline
\hline
\textbf{AH Sco} & U & B & V & R & I & J & H & Ks & L & 12 & 25 & 60 & 100 \\
$\mathrm{\lambda}$ ($\mu$m) & 0.36 & 0.45 &  0.55 & 0.7 & 0.9 & 1.235 & 1.662 & 2.159 & 2.4 & - & - & - & - \\
mag & - & 10.03 & 7.46 & 5.37 & 2.55 & 1.88 & 1.008 & 0.415 & - & - & - & - & - \\
Flux (10$^{-11}$ W / m$^{2}$ $\mu$m) & - & 0.65 & 7.42 & 12.5 & 79.6 & 55.02 & 44.8 & 29.2 & - & 1.31 & 0.168 & 0.006 & 0.001 \\
\hline
\textbf{UY Sct} & U & B & V & R & I & J & H & Ks & L & 12 & 25 & 60 & 100 \\
$\mathrm{\lambda}$ ($\mu$m) & 0.36 & 0.45 &  0.55 & 0.7 & 0.9 & 1.235 & 1.662 & 2.159 & 2.4 & - & - & - & - \\
mag & 14.89 & 11.6 & 9.00 & 6.60 & 4.04 & 2.34 & 1.24 & 0.73 & 0.28 & - & - & - & - \\
Flux (10$^{-11}$ W / m$^{2}$ $\mu$m & 0.0046 & 0.14 & 0.88 & 4.03 & 20.5 & 36.1 & 36.2 & 21.9 & 6.26 & 0.54 & 0.093 & 0.003 & 0.008 \\
\hline
\textbf{KW Sgr} & U & B & V & R & I & J & H & Ks & L & 12 & 25 & 60 & 100 \\
$\mathrm{\lambda}$ ($\mu$m) & 0.36 & 0.45 &  0.55 & 0.7 & 0.9 & 1.235 & 1.662 & 2.159 & 2.4 & - & - & - & - \\
mag & 14.66 & 11.45 & 8.98 & 6.40 & 3.82 & 2.83 & 1.77 & 1.27 & 0.91 & - & - & - & - \\
Flux (10$^{-11}$ W / m$^{2}$ $\mu$m & 0.0057 & 0.18 & 0.89 & 4.85 & 24.7 & 22.98 & 22.15 & 13.27 & 3.50 & 0.52 & 0.07 & 0.0015 & 0.009 \\
\hline
\end{tabular}
\tablefoot{The broadband photometry was obtained from Humphreys \& Ney (\cite{Hum1974}), Kharchenko (\cite{Kharchenko2001}), Cutri et al (\cite{2mass}), and 
IRAS (\cite{iras}). Zero-magnitude flux values are from Skinner (\cite{Skinner1996}) for $UBVRI$ filter and from Cohen et al. (\cite{Cohen2003}) for $JHK$ filter.
The flux values given in this table are not dereddened.}
\label{phot_AHSco}
\end{table*}

We estimated fundamental stellar properties of our sources, in particular
the effective temperature and luminosity, to place them
on the HR diagram and to compare them to stellar evolutionary
tracks. The estimate of the effective temperature was based on 
our interferometric measurement of the angular diameter and
the bolometric flux. The luminosity was derived based on
the bolometric flux and the adopted distance. 

Previous observations of RSGs have shown that continuum interferometric data near 
wavelengths of 2.20\,$\mu$m are represented well by simple limb-darkened disks as predicted by 
hydrostatic model atmospheres (Perrin et al. \cite{Perrin2004}; Ohnaka et al. \cite{Ohnaka2009,Ohnaka2011};  Wittkowski et al. \cite{Witt2012}). Haubois et al. (\cite{Haubois2009}) obtained similar results in the $H$-band. Here, we fitted our continuum data at wavelengths of 2.15-2.25\,$\mu$m to our new 
PHOENIX model atmospheres and derived the angular diameter corresponding
to the model layer where the Rosseland optical depth equals 2/3
(the ``Rosseland angular diameter''). We used a scaled visibility 
function of the form 
\begin{equation}
V(A,\theta_\mathrm{Ross})= 
A \ast V^\mathrm{PHOENIX}(\theta_\mathrm{Ross}),
\end{equation}
where $V^\mathrm{PHOENIX}(\theta_\mathrm{Ross})$ is the synthetic 
visibility function based on the PHOENIX model with Rosseland
angular diameter $\theta_\mathrm{Ross})$ alone, and where $A$
allows the attribution of a fraction of the flux to an over-resolved
dust component.

We used initial parameters of the PHOENIX model based on the 
effective temperature values by van Loon et al. (\cite{VanLoon2005}) 
of T$_\mathrm{eff}$=3574\,K for AH~Sco and UY~Sct and of 
T$_\mathrm{eff}$=3895\,K for KW~Sgr. We 
adopted a surface gravity of log(g)=-0.5 for red supergiants. With these models, we obtained the Rosseland angular radius. Then, we derived the fundamental parameters using the Rosseland angular radius, the bolometric flux, and the distance (as described below in this section). After that, we derived the surface gravity log(g) from the linear Rosseland radius and the assumed stellar mass (estimated from the position of the stars in the HR diagram, 7-10\,M$_{\odot}$). Then, we obtained a new model using the effective temperature and log(g) of our grid closest to the derived values (keeping the chemical composition, micro-turbulent velocity, and stellar mass fixed). We iterated it until the effective temperature and log(g) converged. Finally, we used PHOENIX model atmospheres
with parameters $T_\mathrm{eff}$=3600\,K, $\log(g)=-0.5$
for AH~Sco; $T_\mathrm{eff}$=3400\,K, $\log(g)=-0.5$ for UY~Sct; and $T_\mathrm{eff}$=3700\,K, 
$\log(g)=0.0$ for KW~Sgr. We chose $M$=20\,M$_{\odot}$. Our current model grid includes only masses 
of 1\,M$_{\odot}$ and 20\,M$_{\odot}$, since the structure of the atmospheres is not very sensitive to mass
(Hauschildt et al. \cite{Hausch1999-2}), and the differences in our final parameters based
on 1\,M$_{\odot}$ and 20\,M$_{\odot}$ models are well within the errors. 

We obtained best-fit Rosseland angular diameter values
of $\theta _\mathrm{Ross}$=(5.81~$\pm$~0.15)\,mas, (5.48~$\pm$~0.10)\,mas, and
(3.91~$\pm$~0.25)\,mas for AH~Sco, UY~Sct, and KW~Sgr, respectively,
together with scale factors $A$ of 0.81, 0.94, and 1.0.
For KW~Sgr, we initially obtained a scale factor of 1.08, which
we reset to 1.0, since a factor greater than unity is not physically
possible owing to a dust component.
Using a scaled UD model instead of the PHOENIX model atmospheres, 
we obtained UD angular diameter values of 
$\theta _\mathrm{UD}$=(5.64~$\pm$~0.15)\,mas for AH~Sco, (5.32~$\pm$~0.10)\,mas 
for UY~Sct, and (3.82~$\pm$~0.25)\,mas for KW~Sgr. 
The errors include the statistical errors and systematic errors
caused by calibration uncertainties. We estimated these errors from the differences 
between the visibility curves lying at the maximum and minimum of our data.

\begin{figure}
\centering
\includegraphics[width=1\hsize]{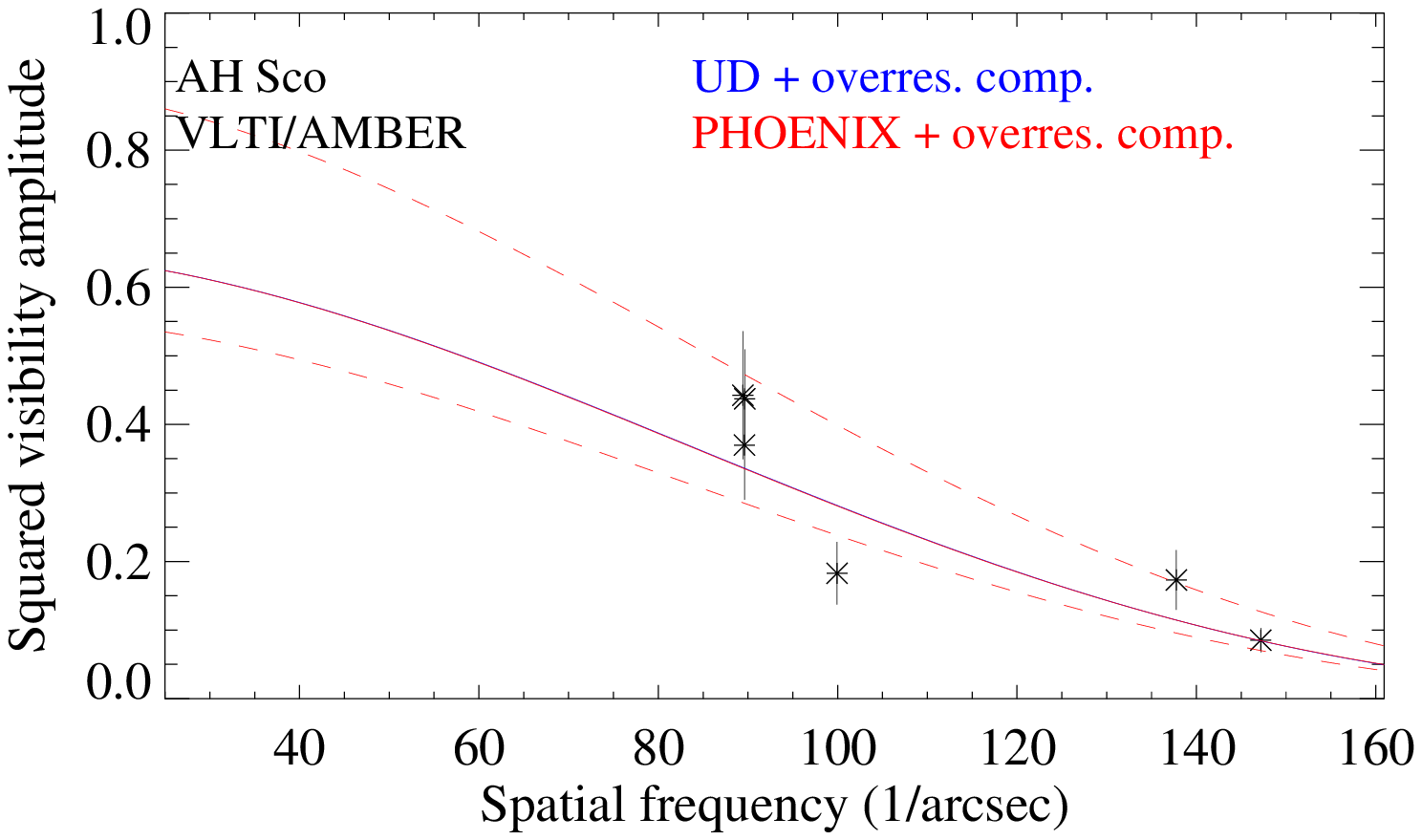}
\includegraphics[width=1\hsize]{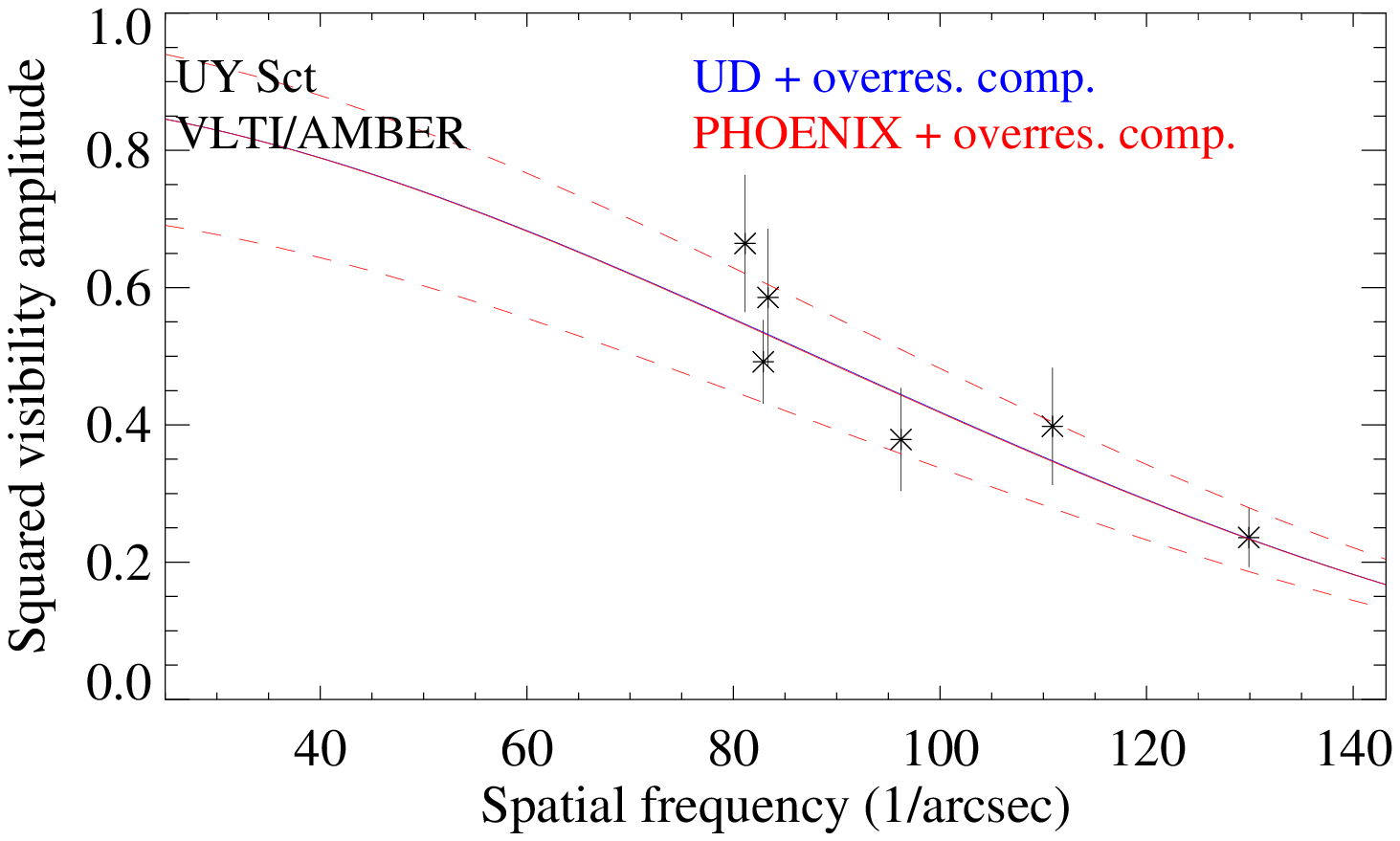}
\includegraphics[width=1\hsize]{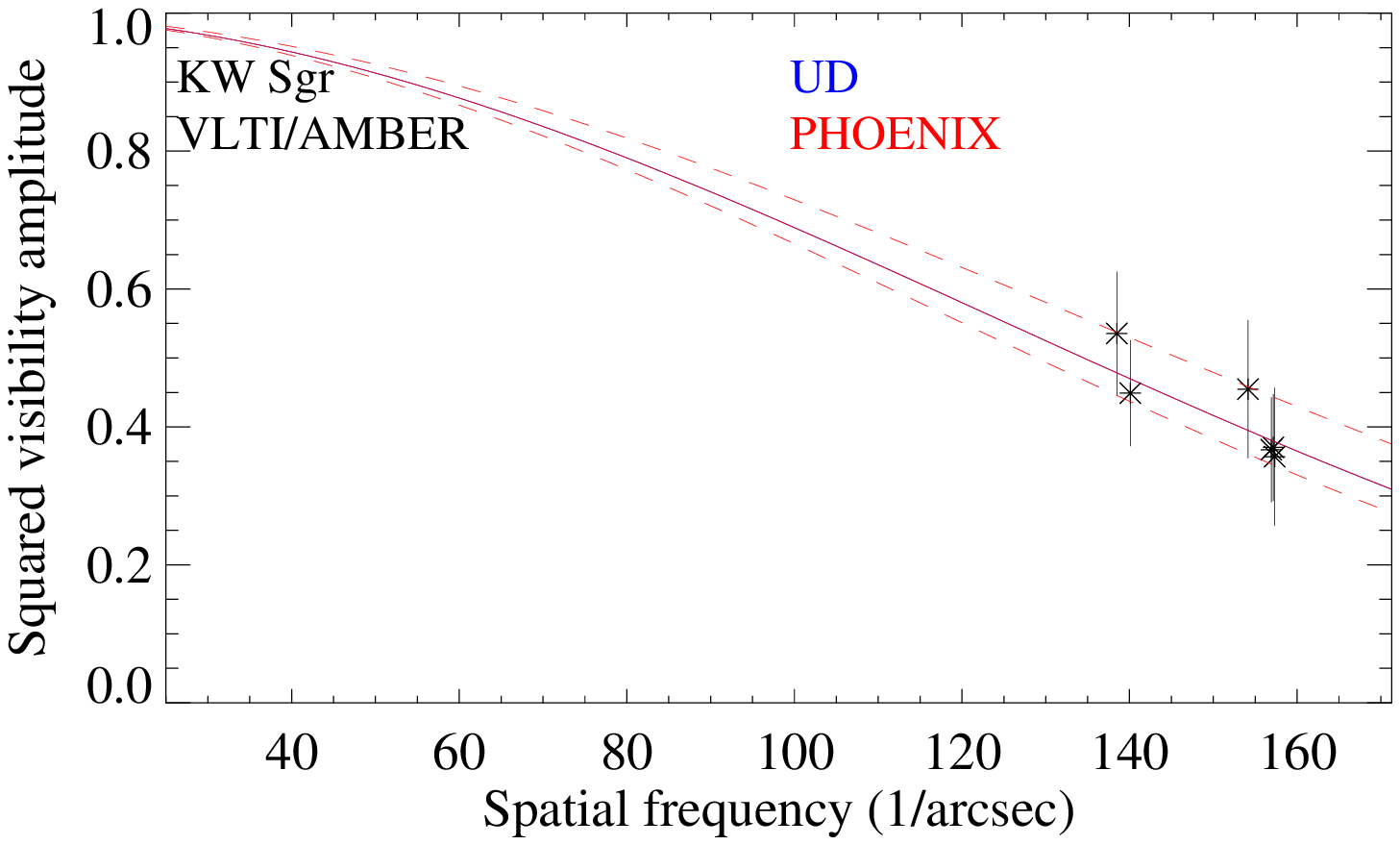}
\caption{Average of squared visibility amplitudes taken in the near-continuum bandpass
at 2.15-2.25\,$\mu$m (one average per baseline and band) for AH~Sco (top), UY~Sct (middle), 
and KW~Sgr (bottom) as a function of
spatial frequency. The red lines indicate the best-fit UD models
and the blue lines the best-fit PHOENIX models. The dashed lines are the maximum and 
minimum visibility curves, from which we estimated the errors of the angular diameters.}
\label{Vis_spacialFrec}
\end{figure}

Figure \ref{Vis_spacialFrec} shows the average of continuum visibility data as a function 
of spatial frequency (one average per baseline and band). Here, the drawn error of the visibility data was computed as an average of the individual errors, since the errors are dominated
by systematic effects. However, the model fit was based directly on the
individual data points. Also shown are the maximum and minimum 
curves corresponding to our adopted error of the angular diameter. The synthetic flux and 
visibility values of the best-fit models are 
indicated in Figs.~\ref{resul_AHSco_fit}, \ref{resul_UYSct_fit},
and \ref{resul_KWSgr_fit}, together with the observed values and with
the best-fit UD model, in all wavelengths. 

To obtain the bolometric flux of our targets, we used $URI$
magnitudes from Humphreys \& Ney (\cite{Hum1974}), $BV$ magnitudes
from Kharchenko (\cite{Kharchenko2001}), $JHK$ magnitudes from 
Cutri et al (\cite{2mass}), and the IRAS fluxes (IRAS \cite{iras}).
We converted the magnitudes into fluxes using 
the zero values from the Skinner (\cite{Skinner1996})
and the 2MASS (Cohen et al. \cite{Cohen2003}) system.

We dereddened the flux values using an estimated $E_{B-V}$
value based on the $V-K$ color excess method.
KW~Sgr belongs to the OB5 Sgr association. Here, we used the
B stars of this association to calculate a mean $E_{B-V}$ value.
AH~Sco and UY~Sct are not known to belong to a cluster,
so that we estimated $E_{B-V}$ using their own magnitudes. 
We used the ratios 
$A_K/A_V=0.12$, $A_V/E_{B-V}=3.69$
for the M stars and 
$A_V/E_{B-V}=3.1$ for the B stars from 
Fiorucci \& Munari (\cite{Fiorucci2003}). 
We estimated the de-reddened value ($A_{\lambda}$) using
$A_\lambda =(\alpha _{\lambda} + 
\beta _{\lambda}\ast E_{B-V})\ast E_{B-V}$, 
where $\alpha _{\lambda}$ and $\beta _{\lambda}$ are taken
for each filter from Morol (\cite{Morol2000}).
We obtained $E_{B-V}$ values of 0.54 mag for AH Sco, 
0.92\,mag for UY~Sct, and 0.98\,mag for KW\,Sgr. For comparison, Levesque et al. (\cite{Levesque2005}) estimate $E_{B-V}$=0.93$\pm$0.33 for KW~Sgr. 
Finally, we integrated the spectro-photometric values to obtain
the bolometric flux. 

It is difficult to find information on the variability of the bolometric flux in RSGs. We (conservatively) estimated the variability of the
bolometric flux using an extrapolation of the relationship obtained by Whitelock et al. (\cite{Whitelock2000}) related to the Hipparcos variability, although this relationship was derived from AGB stars, mostly Mira stars, with larger variability. The final estimated errors are of 15\% for UY~Sct and KW~Sgr, and of 20\% for AH~Sco. The effective temperature was estimated based on the Rosseland angular diameter and the bolometric flux.

The adopted distances to our targets are taken from Melnik \& Dambis
(\cite{Melnik2009}) for KW Sgr and from Chen \& Shen (\cite{Chen2008})
for AH Sco. For UY Sct we used the mean of the distances obtained
by Sylvester et al. (\cite{Sylvester1998}) and the distance obtained with the rule
of Jura et al. (\cite{Jura1989}; A$_K \sim$ 0.15 per kpc distance). 

\begin{figure}
\centering
\includegraphics[width=1\hsize]{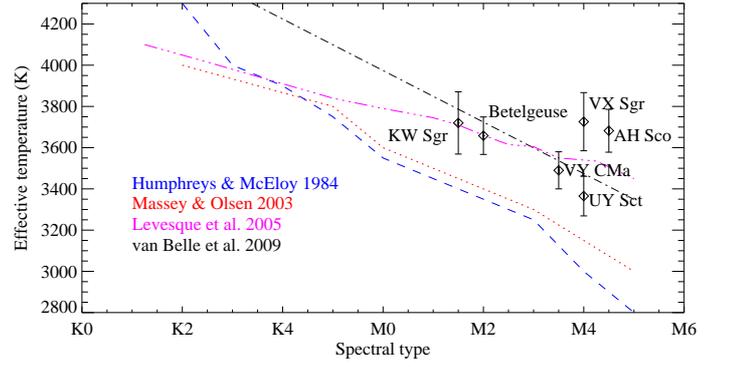}
\caption{Effective temperature versus spectral type of our sources and VY~CMa, VX~Sgr and Betelgeuse. Also we show the scales of Humphreys \& McElroy (\cite{Hum1984}), Massey \& Olsen (\cite{Massey2003}), Levesque et al. (\cite{Levesque2005}), and van Belle et al. (\cite{Belle2009})}
\label{Teff_sp}
\end{figure}

Table \ref{parameters} summarizes the fundamental parameters for 
our targets. The resulting effective temperatures are
T$_\mathrm{eff}$=3682$\pm$190\,K for AH~Sco, 3365$\pm$134\,K for UY~Sct, 
and 3720$\pm$183\,K for KW~Sgr. Figure \ref{Teff_sp} shows the effective temperatures of our targets
versus their spectral types as listed in Table \ref{Log_obs}.
Also shown are three additional RSGs that have
been observed in a similar way using the VLTI/AMBER
instrument, which are VX Sgr (Chiavassa et al. \cite{Chiavassa2010}),
Betelgeuse (Ohnaka et al. \cite{Ohnaka2009,Ohnaka2011}), and VY CMa
(Wittkowski et al. \cite{Witt2012}).
We calculated the effective temperature
and luminosity of Betelgeuse using the angular diameter
obtained by Ohnaka et al. ($\theta _\mathrm{LD}$=42.49\,$\pm$\,0.06\,mas), the distance from Harper et al. (\cite{Harper2008}) (d=197\,$\pm$\,47\,pc),
and the bolometric flux calculated by us in the same
way as for our program stars ($f_\mathrm{bol}$=(1.07\,$\pm$\,0.16)~10$^{-10}$\,W~m$^{2}$). 
We estimated the effective temperature to be 3620$\pm$137~K. This value is consistent 
with the estimated by Perrin et al. (\cite{Perrin2004}) of 3641$\pm$53~K, Haubois et al. (\cite{Haubois2009})
of 3600$\pm$66~K, and Ohnaka et al. (\cite{Ohnaka2011}) of 3690$\pm$54~K. Likewise, we 
calculated the properties of VX Sgr
using the photospheric angular diameter of
8.82\,$\pm$\,0.5\,mas by Chiavassa et al. (\cite{Chiavassa2010}), the distance
of 1570\,$\pm$\,270\,pc by Chen et al. (\cite{Chen2007}), and the bolometric
flux calculated by us ($f_\mathrm{bol}$=(5.00\,$\pm$\,1.0)~10$^{-9}$\,W~m$^{2}$).

\begin{table}
\caption{Fundamental parameters of AH~Sco, UY~Sct, and KW~Sgr}
\centering
\begin{tabular}{lcccc}
\hline
\hline
 Parameter & AH Sco & UY Sct & KW Sgr & Ref. \\
 \hline
  f$_\mathrm{bol}$ & 2.07 $\pm$ 0.41 & 1.28 $\pm$ 0.19 & 0.97 $\pm$ 0.15 & 1\\ 
  ($10^{-9}$ W m$^{-2}$) & & & & \\
  d (pc) & 2260$\pm$ 190 & 2900 $\pm$ 317 & 2400 $\pm$ 300 & 2\\
  L ($10^{32}$ W) & 1.26 $\pm$ 0.33 & 1.29 $\pm$ 0.34 & 0.67 $\pm$ 0.20 & 3\\
  log(L/L$_{\odot}$) & 5.52 $\pm$ 0.26 & 5.53 $\pm$ 0.27 & 5.24 $\pm$ 0.29 & -\\
  $\theta _\mathrm{Ross}$ (mas) & 5.81 $\pm$ 0.15 & 5.48 $\pm$ 0.10 & 3.91 $\pm$ 0.25 & 4\\
  R(R$_\odot$) & 1411 $\pm$ 124 & 1708 $\pm$ 192 & 1009 $\pm$ 142 & 5\\
  T$_\mathrm{eff}$ (\,K) & 3682 $\pm$ 190 & 3365 $\pm$ 134 & 3720 $\pm$ 183 & 6\\
  log(T$_\mathrm{eff}$) & 3.57 $\pm$ 0.05 & 3.53 $\pm$ 0.04 & 3.57 $\pm$ 0.05 & -\\
  \hline
\end{tabular}
\tablefoot{1: Humphreys \& Ney (\cite{Hum1974}), Kharchenko (\cite{Kharchenko2001}), 
Cutri et al (\cite{2mass}), IRAS (\cite{iras}). 
2: Chen  \& Shen \cite{Chen2008}. (AH~Sco), 
Sylvester et al. \cite{Sylvester1998} (UY~Sct), 
Melnik, A. M., Dambis, A. K., \cite{Melnik2009}, (KW~Sgr). 
3: from 2 and 1; 
4: This work; 
5: from 4 and 2; 
6: from 5 and 3. }
\label{parameters}
\end{table}

It has been discussed that VX Sgr may be either
a relatively massive AGB star or a red supergiant
(Chiavassa et al. \cite{Chiavassa2010}).
Our values suggest that VX Sgr is an
RSG, so we used the spectral type of
M4\,Ia by Humphreys et al. (\cite{Hum1972}) instead
of the Simbad classification of M5/M6 III
by Houk \&  Smith-Moore (\cite{Houk1988}).
For comparison, Fig. \ref{Teff_sp} includes the calibrations
of the effective temperature scales by Humphreys \& McEloy
(\cite{Hum1984}), Massey \& Olsen (\cite{Massey2003}) , Levesque et al. (\cite{Levesque2005}), and van Belle et al. (\cite{Belle2009}).
Our values confirm the more recent calibrations
by Levesque et al. and van Belle et al.
In particular, KW Sgr and Betelgeuse are in
very good agreement with the scale by Levesque
et al., while the cooler sources show a larger
scatter, but are consistent with both of these
calibrations within 1-2 $\sigma$. This scatter may be caused by an erroneous 
calibration, and also conceivable variations of the spectral type (Humphreys et al. \cite{Hum1972}). 
 
\begin{figure}
\centering
\includegraphics[width=1\hsize]{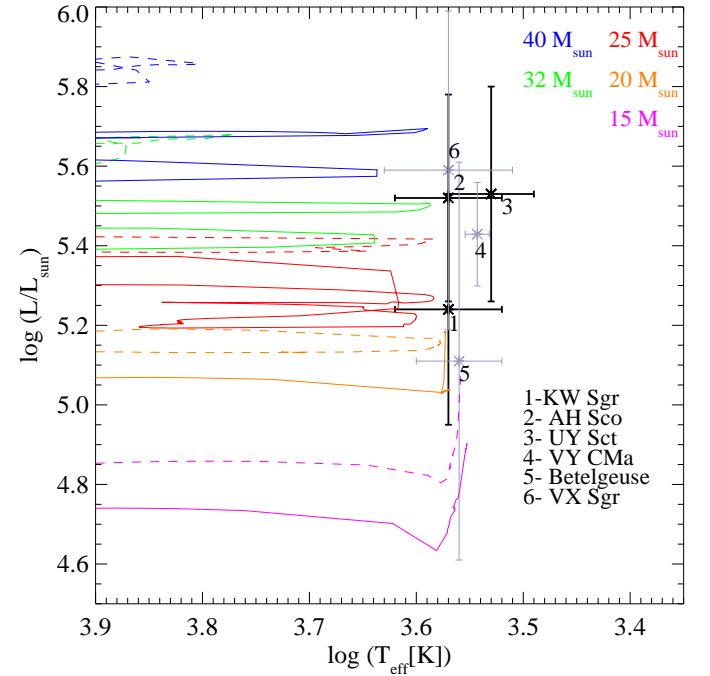}
\caption{Location of AH~Sco, UY~Sct, and KW~Sgr
in the HR diagram using our fundamental parameters.
Also shown is the position of VY~CMa from Wittkowski et al. (\cite{Witt2012}), Betelgeuse from Ohnaka et al. (\cite{Ohnaka2011}), and VX~Sgr from (Chen et al. \cite{Chen2007} and Chiavassa et al.
(\cite{Chiavassa2010}).
The positions of the stars are comopared to evolutionary tracks from
Ekstr\"om et al. (\cite{Ekstrom2012}) for masses of 15\,M$_{\odot}$,
20\,M$_{\odot}$, 25\,M$_{\odot}$, 32\,M$_{\odot}$, and 40\,M$_{\odot}$.
The solid lines are models without rotation, and the dashed lines are with rotation.}
\label{diagram_HR}
\end{figure}

Figure \ref{diagram_HR} shows the positions of AH~Sco, UY~Sct, and KW~Sgr in 
the Hertzsprung-Russell (HR) diagram, together with the recent evolutionary 
tracks from Ekstr\"om et al. (\cite{Ekstrom2012}). Also shown are the positions of VY~CMa, Betelgeuse, and VX~Sgr as discussed above.
The targets are located close to the red limits of 
these tracks. AH Sco and UY Sct are close to the evolutionary tracks corresponding to an initial mass of 25\,M$_{\odot}$ with rotation or 30-40\,M$_{\odot}$ without rotation. KW Sgr is consistent with evolutionary tracks corresponding to masses ranging from 20\,M$_{\odot}$ with or without rotation, to 40\,M$_{\odot}$ without rotation. The positions of our targets with higher mass ($\geq$~20~M$_{\odot}$) are drawn slightly to the right of 
the Hayashi limit. This may be consistent with Fig. 3c of Levesque et al. (\cite{Levesque2005}), where several sources of this mass are placed slightly to the right, too. This may indicate that sources of this mass are not in hydrostatic equilibrium. Our measurements are based on direct measurements of the angular
diameter and the bolometric flux, and thus provide an independent
determination of the location of red supergiants in the HR diagram
compared to the approach by Levesque et al. (\cite{Levesque2005}), which
was based on a comparison of spectro-photometry to hydrostatic MARCS
model atmospheres.


\section{Atmospheric structure}

Previous studies have demonstrated that near-infrared spectra of RSGs are 
predicted reasonably well by static PHOENIX model atmospheres 
(Lan\c con et al. \cite{Lancon2007}), but that discrepancies are observed
in detailed spectral features toward the coolest stars. 
Lan\c con et al. noted that these discrepancies may be related to the 
variability of these sources and the higher micro-turbulent velocities.

The top panels of Figs.~\ref{resul_AHSco_fit}-\ref{resul_KWSgr_fit}
indicate the flux spectra of the best-fit PHOENIX model atmospheres
compared to the observed flux spectra.
They show a good agreement of the overall shapes of the 
1.9-2.5\,$\mu$m spectra, and in particular good agreement
of the strengths of the CO bandheads. The largest deviations are observed
for the spectral shape of the very cool M4 supergiant UY~Sct.
Overall, our results on the $K$-band spectra are consistent
with those by Lan\c con et al. (\cite{Lancon2007}).

The synthetic interferometric visibility spectra of the best-fit 
PHOENIX models are compared to the observed visibility spectra in the 
second panels of Figs.~\ref{resul_AHSco_fit}-\ref{resul_KWSgr_fit}. 
Considering the errors bars, the continuum visibility values are consistent 
with a limb-darkened disk. This comparison shows that the large observed 
drops of the visibility function 
at the locations of the CO bandheads between 2.3\,$\mu$m and 2.5\,$\mu$m,
as well as of the broad H$_2$O feature centered on 1.9\,$\mu$m
are not reproduced by the PHOENIX model atmospheres that we used here.
The synthetic visibility spectra show features at the location
of the CO bandheads, but which are much weaker than the 
observed features.
The observed drops of the visibility at the H$_2$O feature centered 
on 1.9\,$\mu$m are not visible in the model visibility spectra.

These results indicate that the opacities of CO are included in the 
PHOENIX models providing a satisfactory fit to the $K$-band spectra,
but that the extension of these molecular layers are too compact in 
the models compared to our observations. This result is consistent with 
the similar observations of the RSG VY CMa by 
Wittkowski et al. (\cite{Witt2012}) and the results shown by Lan\c con et al. (\cite{Lancon2007}).
The satisfactory fit to stellar spectra, together with an unsatisfactory
fit to interferometric visibilities, illustrates the importance of 
interferometric observations in addition to spectral observations 
to fully constrain and test stellar model atmospheres, as noted 
previously by, e.g.,  Wittkowski et al. (\cite{Witt2006}) and 
Paladini et al. (\cite{Paladini2011}). In a future work, it may be possible to extend PHOENIX model atmospheres
farther out, for instance by attaching a slow wind to the
atmosphere, or by considering NLTE in the molecular lines
or 3D structures.

The levitation of gas to radii where dust particles can condense
and be accelerated by radiative pressure, as observed here, is an 
important ingredient in our understanding of the formation of 
stellar winds of cool giant and supergiant stars
(e.g. Bladh \& H\"ofner \cite{Bladh2012}). It is not yet
clear which physical processes may levitate the gas in the case of 
supergiant stars.
In the case of asymptotic giant branch (AGB) stars, it has been
established that pulsation-induced shock waves lead to very extended
atmospheres that reach to radii of a few photospheric radii 
(Jeong et al. \cite{Jeong2003}; Ireland et al. \cite{Ireland2004,Ireland2008,Ireland2011}).
Recent interferometric observations of Mira variable AGB stars 
obtained with the AMBER instrument by Wittkowski et al. (\cite{Witt2011})
have shown extended molecular layers that are consistent with 
predictions by the latest hydrodynamical model atmospheres of 
Mira variables (Ireland et al. \cite{Ireland2011}). 

It is remarkable that 
the wavelength dependence of the visibility spectra observed here for 
RSGs is qualitatively similar to observations of AGB stars (Marti-Vidal et al. \cite{Marti2011}; Ruiz-Velasco et al. \cite{Ruiz-Velasco2011}; Wittkowski et al. \cite{Witt2011}) and to 
predictions by the latest hydrodynamic model atmospheres of Miras. 
This has also already been observed for VX~Sgr (Chiavassa et al. \cite{Chiavassa2010}). 
Although the pulsation amplitudes of RSGs are generally lower than those 
of Mira variable AGB stars, we speculate that the same process of 
pulsation-induced shock waves may also lead to the extended molecular 
layers that we observed here for the RSGs AH~Sco, UY~Sct, and KW~Sgr, and 
that have also been observed previously for other RSGs. 
Samus et al. (\cite{samus}) 
list V-band amplitudes of 2-4 mag and pulsation periods around 700 days
for AH~Sco, KW~Sgr, and UY~Sct, compared to 4-6~mag and periods between
300-450 days for the Mira variables discussed in 
Wittkowski et al. (\cite{Witt2011}). Indeed, Heger et al. (\cite{Heger1997}) found that RSGs can have similar pulsation properties as AGB stars, based on their theoretical calculation. Dynamic model atmospheres based on 
stellar pulsation are not currently available for the parameter space of RSGs, 
so that this hypothesis cannot be tested within this paper. We encourage the
development of such dynamic model atmospheres for RSGs.

SiO maser emission has been observed to form in ring-like structures
around a large number of oxygen-rich evolved stars (cf., Benson et al.
\cite{Benson1990}). In the past, radii of the SiO maser emission were compared
to stellar radii to obtain an additional constraint on the structure of the
extended atmosphere and on the maser pumping mechanism. Most of these
comparisons have been obtained for Mira variable stars. When meaningful
stellar radii such as continuum radii or Rosseland radii were used and
were obtained at similar phases as the SiO maps, the radii of the maser
emission was consistently found close to two stellar radii
(e.g., Boboltz \& Wittkowski \cite{Boboltz2005};
Fedele et al. \cite{Fedele2005}; Cotton et al. \cite{Cotton2010}).
This location is consistent with theoretical modeling by
Humphreys et al. (\cite{Humphreys1997}) and Gray et al. (\cite{Gray2009}),
and is predicted to be located close to the position of a shock front.
Among the RSGs observed here, AH Sco is also a known source of SiO
maser emission with a mean diameter of 17.2\,mas
(Chen \& Shen \cite{Chen2008}).
Compared to our Rosseland angular diameter of 6.0 mas, the maser emission
lies at 2.9 stellar radii. Similarly, the maser radius of VX Sgr of 13 mas
by Su et al. (\cite{Su2012}) lies at 3.0 stellar radii compared to the
Rosseland angular diameter of 8.8 mas by Chiavassa et al.
(\cite{Chiavassa2010}). This might indicate that SiO maser of red supergiants
are located at systematically larger distances from the stellar surface
compared to AGB stars. Within a framework of stellar pulsation,
this may be related to the larger pulsation periods and a different
shock front behavior. 
  
Convection has also been discussed as a possible process to dredge material up
to higher layers and to produce surface inhomogeneities and shock
waves in the photosphere (e.g., Chiavassa et al. \cite{Chiavassa2011}).
Available 3D convection simulations for RSGs have been compared
to interferometric data of VX Sgr and have not shown the typical observed
wavelength dependence of the visibility function at the positions
of the water band centered at 1.9\,$\mu$m and of the CO bandheads
at 2.3\,$\mu$m-2.5\,$\mu$m (Chiavassa et al. \cite{Chiavassa2010}). 
Ohnaka et al. (\cite{Ohnaka2011}) discuss
that the observed time variation of the continuum data of the RSG
Betelgeuse is much less than the maximum variation predicted by
current 3D convection simulations, suggesting that the effect of
photospheric convection may be overestimated in current models.
Chiavassa et al. (\cite{Chiavassa2011}) mention that with current
models it is not clear whether convective motion is strong enough
to explain the motion detected at extended layers up to 1.4 stellar
radii, where the convective flux is low. While these results suggest
that the effects of convection may be limited to the photospheric
layers to a smaller degree than present in current models and may not 
reach into extended molecular layers, a detailed comparison of
interferometric data to latest 3D convection simulations is required
for a definite answer, and will be a part of our future work.


\section{Conclusions}

We obtained near-infrared spectro-interferometric observations of the RSGs
AH Sco, UY Sct, and KW Sgr. Our data indicate the presence of extended
molecular layers of CO and H$_2$O. We compared the spectra and visibilities
to a new grid of hydrostatic PHOENIX model atmospheres.

Our near-infrared spectra of AH~Sco, UY~Sct, and KW~Sgr are reproduced well 
by the PHOENIX model atmospheres; however, the synthetic visibility 
amplitudes of the model do not predict the large extensions of the molecular 
bands. This may be due to processes including pulsation or convection, which are not included in our currently available hydrostatic model atmospheres. These processes may explain the observed extensions of the atmospheres and the wind acceleration in RSGs.
 
The continuum (2.15-2.25\,$\mu$m) 
appears to be largely free of contamination by molecular layers. Thus, the PHOENIX model fitting to the continuum allows us to estimate meaningful stellar radii. 
Using the PHOENIX model in the continuum, we estimate a Rosseland diameter 
of 5.81$\pm$0.15\,mas (R=1411$\pm$124\,R$_{\odot}$) for AH~Sco, 
5.48$\pm$0.10\,mas (R=1708$\pm$192\,R$_{\odot}$) for UY~Sct, and 
3.91$\pm$0.25\,mas (R=1009$\pm$142\,R$_{\odot}$) for KW~Sgr. We also estimate
effective temperatures of 3682$\pm$190\,K for AH~Sco, 3365$\pm$134\,K for 
UY~Sct, and 3720$\pm$183\,K for KW~Sgr.

With the effective temperature and the luminosity calculated from the 
Rosseland radius, the bolometric flux, and the distance, we locate our targets 
in the HR diagram. The positions of the stars 
in the HR diagram fall close to the red limit of the tracks corresponding 
to stars of mass around  25/30-40~M$_{\odot}$ (AH~Sco and UY~Sct) and 20/20-40~M$_{\odot}$ (KW~Sgr) with/without rotation. 

Spectro-interferometric observations of RSGs with spatially resolved extended molecular layers are still scarce. In this work, we studied three sources, which almost doubles the sample size.
 
\begin{acknowledgements}
We gratefully thank Juan Fabregat and Gerard van Belle for
their great help with the calculation of the extinction values and the
bolometric fluxes. This research made use of the AMBER data
reduction package of the Jean-Marie Mariotti Center, and the SIMBAD
database operated at the CDS, Strasbourg, France. B.A.T and J.M.M acknowledge
support by the Spanish Ministry of Science and Innovation though the grant
AYA2009-13036-C02-02.

\end{acknowledgements}


\begin{thebibliography}{}

   \bibitem[1990]{Benson1990} Benson, P.~J., Little-Marenin, I.~R. \&
Woods, T.~C., et al.\ 1990, \apjs, 74, 911

   \bibitem[1998]{Bessel1998} Bessel, M. S., Castelli F., \& Plez, B., 1998, A\&A, 333, 231

   \bibitem[2012]{Bladh2012} Bladh, S., \& H\"ofner, S. 2012, A\&A, 546, A76
   


	\bibitem[2005]{Boboltz2005} Boboltz, D.~A., \& Wittkowski, M.\ 2005,
\apj, 618, 953
   
   \bibitem[1985]{Campins1985} Campins, H., Rieke, G. H., \& Lebofsky, M. J., 1985, AJ, 90, 896

   \bibitem[2009]{Chelli2009} Chelli, A., Utrera, O. H., \& Duvert, G. 2009, A\&A, 502, 705
   
   \bibitem[2007]{Chen2007} Chen, X., Shen, Z.-Q., \& Xu, Y.\ 2007, \cjaa, 7, 531 
   
   \bibitem[2008]{Chen2008} Chen, X., \& Shen, Z.-Q., 2008,  ApJ, 681, 1574
   
   \bibitem[2010]{Chiavassa2010} Chiavassa, A., Lacour, S., Millour, F., et al. 2010, A\&A, 511, A51
 
   \bibitem[2011]{Chiavassa2011} Chiavassa, A., Freytag, B., Masseron, T. et al, 2011, A\&A, 535, A22 
   
   \bibitem[2003]{Cohen2003} Cohen, M., Wheaton, Wm. A., \& Megeath, S. T., 2003, AJ, 126, 1090
     
	\bibitem[2010]{Cotton2010} Cotton, W.~D., Ragland, S., Pluzhnik, E.~A.,
et al.\ 2010, \apjs, 188, 506
   
   	\bibitem[2003]{2mass} Cutri, R.~M., Skrutskie, M.~F., van Dyk, S., et al.\ 2003, ''The IRSA 2MASS All-Sky Point Source Catalog, NASA/IPAC Infrared Science Archive.

   \bibitem[2012]{Ekstrom2012} Ekstrom, S., Georgy, C., Eggenberger, P., et al. 2012, A\&A, 537, 146
   
   \bibitem[2005]{Fedele2005} Fedele, D., Wittkowski, M., \&
Paresce, F., et al.\ 2005, \aap, 431, 1019
   
   \bibitem[2003]{Fiorucci2003} Fiorucci, M., \& Munari, U., 2003, A\&A, 401, 781  
   
   \bibitem[2008]{Harper2008} Harper, G.~M., Brown, A., \& Guinan, E.~F.\ 2008, \aj, 135, 1430 

   \bibitem[2009]{Haubois2009} Haubois, X., Perrin, G., Lacour, S., et al. 2009, A\&A, 508, 923
   
   \bibitem[1999]{Hausch1999-2} Hauschildt, P. H., Allard, F., Ferguson, J., Baron, E., \& Alexander, D. R. 1999, ApJ, 525, 871

   \bibitem[1999]{Hausch1999} Hauschildt, P. H., \& Baron , E. 1999, J. Comp. Appl. Math, 109, 41

   
   \bibitem[1997]{Heger1997} Heger, A., Jeannin, L., Langer, N., \& Basaffe, I, 1997, A\&A, 327, 224
   
   \bibitem[1988]{Houk1988} Houk, N., \& Smith-Moore, M.\ 1988, Michigan Catalogue of Two-dimensional Spectral Types for the HD Stars.~Volume 4

   \bibitem[1972]{Hum1972} Humphreys, R. M., Strecker, D. W., \& Ney, E. P. 1972, ApJ, 172, 75
      
   \bibitem[1974]{Hum1974} Humphreys, R. M., \& Ney, E. P. 1974, ApJ, 194, 623
   
   \bibitem[1984]{Hum1984} Humphreys, R. M., \& McElroy, D. B. 1984, ApJ, 284, 565
   
   \bibitem[1997]{Humphreys1997} Humphreys, E.~M.~L., Gray, M.~D.,
Field, D., Yates, J.~A., \& Bowen, G.\ 1997, \apss, 251, 215
   
   	\bibitem[1986]{iras} IRAS Catalogue of Point Sources (V2.0, 1986)

  \bibitem[2004]{Ireland2004} Ireland, M. J., Scholz, M., \& Wood, P. R. 2004, MNRAS, 352, 318

  \bibitem[2008]{Ireland2008} Ireland, M. J., Scholz, M., \& Wood, P. R. 2008, MNRAS, 391, 1994

  \bibitem[2011]{Ireland2011} Ireland, M. J., Scholz, M., \& Wood, P. R. 2011, MNRAS, 418, 114

   \bibitem[2003]{Jeong2003} Jeong, K. S., Winters, J. M., Le Bertre, T., \& Sedlmayer, E. 2003, A\&A, 407, 191

   \bibitem[1965]{Johnson1965} Johnson, H. L., 1965, CoLPL, 3, 73.

   \bibitem[1989]{Jura1989} Jura, M., Joyce, R., R., \& Kleinmann, S. G., 1989, ApJ, 336, 936
 
   \bibitem[1990]{Jura1990} Jura, M., \& Kleinmann, S. G., 1990, ApJ, 769, 73
   
   \bibitem[2009]{Gray2009} Gray, M. D., Wittkowski, M., et al. 2009, MNRAS, 397, 51
   
   \bibitem[2009]{Kervella2009} Kervella, P., Verhoelst, T., Ridgway, S.~T., et al.\ 2009, \aap, 504, 115

   \bibitem[2011]{Kervella2011} Kervella, P., Perrin, G., Chiavassa, A., et al.\ 2011, \aap, 531, A117
   
   \bibitem[2001]{Kharchenko2001} Kharchenko, N. V., 2001, KFNT, 17, 409   
      
   \bibitem[2010]{Lafrasse2010} Lafrasse, S., Mella, G., Bonneau, D., et al. 2010, Proc. SPIE, 7734, 410
   
   \bibitem[2000]{Lancon2000} Lan\c con, A., \& Wood, P. R. 2000, A\&AS, 146, 217
   
   \bibitem[2007]{Lancon2007} Lan\c con, A., Hauschildt, P. H., Ladjal, D. \& Mouhcine, M., 2007, A\&A, 468, 205
  
  \bibitem[2005]{Levesque2005} Levesque, E. M., Massey, P., Olsen, K. A. G., et al. 2005, ApJ, 628, 973
      
   \bibitem[1992]{Lord1992} Lord, S. D. 1992, NASA technical memorandum 103957

   \bibitem[2011]{Mauron2011} Mauron, N., Josselin, E., 2011, A\&A, 526, A156 
   
   \bibitem[2011]{Marti2011} Mart{\'{\i}}-Vidal, I., Marcaide, J.~M., Quirrenbach, A., et al.\ 2011, A\&A, 529, A115 
   
   \bibitem[2003]{Massey2003} Massey, P., \& Olsen, K. A. G. 2003, ApJ, 126, 2867.

   \bibitem[2006]{Massey2006} Massey, P., Levesque, E. M., \& Plez, B., 2006, ApJ, 646, 1203.

   \bibitem[2009]{Melnik2009} Melnik, A. M, \& Dambis, A. K., 2009, MNRAS, 400,518

   \bibitem[1978]{Morel1978} Morel, M., \& Magnenat, P., 1978, A\&AS, 34, 477 
   
   \bibitem[2000]{Morol2000} Moro1, D., \& Munari1, U. 2000, A\&A, 147, 361
   
   

   \bibitem[1992]{IPAC1992} Moshir M., Kopan G., Conrow T., et al., 1992, Infrared Processing and Analysis Center (IPAC), California Institute of Technology

   \bibitem[2009]{Ohnaka2009} Ohnaka, K., Hofmann, K. H., Benisty, M., et al., 2009, A\&A, 503, 183

   \bibitem[2011]{Ohnaka2011} Ohnaka, K., Weigelt, G., Millour, F., et al., 2011, A\&A, 529, 163

   \bibitem[2011]{Paladini2011} Paladini, C., van Belle, G. T., Aringer, B., et al. 2011, A\&A, 533, A27

   \bibitem[2004]{Perrin2004} Perrin, G., Ridgway, S. T., Coud\'e du Foresto, V., et al., 2004, A\&A, 418, 675.
   
   \bibitem[2005]{Perrin2005} Perrin, G., Ridgway, S. T., Verhoelst, T., et al. 2005, A\&A, 436, 317

  \bibitem[2007]{Petrov2007} Petrov, R. G., Malbet, F., Weigelt, G., et al. 2007, A\&A, 464,1

   \bibitem[1988]{Tarafdar1988} Tarafdar, \& S. P., 1988, ApJ, 331, 932

   \bibitem[2007]{Tatulli2007} Tatulli, E., Millour, F., Chelli, A., et al. 2007, A\&A, 464, 29  

   \bibitem[2006]{Tsuji2006} Tsuji, T. 2006, ApJ, 645, 1448   
   
   \bibitem[2011]{Ruiz-Velasco2011} Ruiz-Velasco, A.~E., Wittkowski, M., Wachter, A., Schr{\"o}der, K.-P., \& Driebe, T.\ 2011, \aap, 535, A100 
   

	\bibitem[2009]{samus} Samus, N.~N., Durlevich, O.~V., \& et al.\ 2009, VizieR Online Data Catalog, 2025
	
	\bibitem[1996]{Skinner1996} Skinner, C. J., 1996, Instrument Science Report NICMOS-014
	
	\bibitem[2012]{Su2012} Su, J.~B., Shen, Z.-Q., \& Chen, X., et al.\ 2012,
\apj, 754, 47 

   \bibitem[1998]{Sylvester1998} Sylvester, R. J., Skinner, C. J., \& Barlow, M. J., MNRAS, 301, 1083

   \bibitem[2009]{Belle2009} van Belle, G. T., Creech-Eakman, M. J., \& Hart, A., 2009, MNRAS, 394, 1925.

   \bibitem[2005]{VanLoon2005} Van Loon, J. Th., Cioni, M. R. L., Zijlstra, A. A. et al. 2005, A\&A, 438, 273

    \bibitem[2000]{Whitelock2000} Whitelock, P., Feast, M., 2000, MNRAS 319, 728 

   \bibitem[2003]{Witt2003} Wittkowski, M., Aufdenberg, J., P, \& Kervella, P. 2003, A\&A, 413, 711

   \bibitem[2006]{Witt2006} Wittkowski, M., Aufdenberg, J. P., Driebe, T., et al. 2006, A\&A, 460, 855
   
   \bibitem[2011]{Witt2011} Wittkowski, M., Boboltz, D. A., Ireland, M., et al. 2011, A\&A, 532L, 7   
     
   \bibitem[2012]{Witt2012} Wittkowski, M., Hauschildt, P. H., Arroyo-Torres, B., \& Marcaide, J. M., 2012, A\&A, 540, 12
   
    
   
    

   



 

 





\end{thebibliography}
\end{document}